%% file: paper.tex
\documentclass[conference]{IEEEtran}
\usepackage[T1]{fontenc}
\usepackage{textcomp}
\usepackage{amsmath}
\usepackage{amssymb}
\usepackage{dsfont}
\usepackage{mleftright}
\usepackage{graphicx}
\usepackage{booktabs}
\usepackage{cite}
\usepackage{url}
\usepackage{algorithm}
\usepackage{algpseudocode}

\algdef{SE}[BLOCK]{Block}{EndBlock}[1]{#1}{}
\algtext*{EndBlock}

\newtheorem{proposition}{Proposition}

\newcommand*{\bigtimes}{\mathop{\mathchoice{\vcenter{\hbox{\LARGE$\times$}}}{\vcenter{\hbox{\large$\times$}}}{\vcenter{\hbox{$\scriptstyle\times$}}}{\vcenter{\hbox{$\scriptscriptstyle\times$}}}}\displaylimits}

\usepackage[hidelinks]{hyperref}
\hypersetup{
  pdftitle={Joint Dynamic Lane Reversal and Variable Speed Limit Control: A Graphormer-Based Approach},
  pdfauthor={Hamed Araab, Abbas Ahmadi, Mohammad Hossein Basiri}
}

\begin{document}
\bstctlcite{IEEEexample:BSTcontrol}

\title{Joint Dynamic Lane Reversal and Variable Speed Limit Control: A Graphormer-Based Approach}

\author{\IEEEauthorblockN{Hamed Araab}
\IEEEauthorblockA{\textit{Department of Industrial Engineering}\\
\textit{and Management Systems}\\
\textit{Amirkabir University of Technology}\\
\textit{(Tehran Polytechnic)}\\
Tehran, Iran\\
hamedaarab@aut.ac.ir}
\and
\IEEEauthorblockN{Abbas Ahmadi}
\IEEEauthorblockA{\textit{Department of Industrial Engineering}\\
\textit{and Management Systems}\\
\textit{Amirkabir University of Technology}\\
\textit{(Tehran Polytechnic)}\\
Tehran, Iran\\
abbas.ahmadi@aut.ac.ir}
\and
\IEEEauthorblockN{Mohammad Hossein Basiri}
\IEEEauthorblockA{\textit{Department of Electrical Engineering}\\
\textit{Amirkabir University of Technology}\\
\textit{(Tehran Polytechnic)}\\
Tehran, Iran\\
basiri@aut.ac.ir}}

\maketitle

\begin{abstract}
\input{sections/abstract}
\end{abstract}


\input{sections/introduction}
\input{sections/prior-work}
\input{sections/problem-formulation}
\input{sections/architecture}
\input{sections/experimental-design}
\input{sections/results}
\input{sections/discussion}
\input{sections/conclusion}
\input{sections/acknowledgments}

\bibliographystyle{IEEEtran}
\bibliography{refs}

\end{document}

%% file: sections/abstract.tex
Connected and automated vehicles could allow reassigning lane directions and speed limits on every road frequently, but the benefit hinges on how traffic reroutes, and one-way roads can isolate junctions. We formulate network-wide dynamic lane allocation and directional speed limit control with one-way roads, lane clearing before each reversal, and a navigable graph kept strongly connected at all times. An action-resolution algorithm grounded in graph orientation theory admits a proposal closing a direction only if the network stays strongly connected and otherwise relaxes or reverts it, keeping all junctions mutually reachable within the model, even during clearing. All road controllers share one transformer whose attention, Global Endpoint Relation-Aware Spatio-Temporal Self-Attention (GERASTSA), spans all road--time tokens, biased by the endpoint relations between roads. It is pretrained on action-conditioned traffic forecasting and fine-tuned with per-road proximal policy optimization under a mixed reward that also credits neighboring roads and the network. On a $6 \times 6$ grid with commuter rush hours, the learned policy makes about two thirds of roads one-way in the peak direction and cuts daily delay by $90 \%$ against the initial configuration held fixed and by $85 \%$ against a tuned two-way load balancer. Restricted to two-way roads at test time, it gains only about as much as this balancer, whereas its learned speed limits add delay. Pretraining roughly halves the updates needed to surpass the fixed configuration. Of the ablations, only removing the relational bias or the spatial embedding clearly matters, multiplying delay about fivefold.

%% file: sections/introduction.tex
\section{Introduction}\label{sec:intro}
Urban traffic demand is directionally unbalanced. In the morning, commuters travel from residential areas toward commercial ones, and in the evening they return, so that the lanes leading in one direction are congested while those leading in the other are lightly used. The lane directions of most roads, however, are fixed. Reversible lanes and contraflow operations relax this restriction by assigning lanes to the direction that needs them \cite{ext:trb2004convertible,ext:wolshon2001contraflow}, and variable speed limits (VSLs) regulate how traffic uses the capacity that a road offers \cite{ext:lu2014vsl}. In practice, both are usually deployed on individual corridors, where reversible lanes typically follow preset schedules and VSLs react to local conditions. Connected and automated vehicles (CAVs) change this picture: they can be informed of a new lane direction or speed limit immediately and, if they comply with it reliably, make it conceivable to reconfigure every road of a network at short intervals in response to current traffic \cite{ext:hausknecht2011dlr,ext:levin2016dlr}.

Reconfiguring a network, rather than a corridor, raises two difficulties. The first is that the effect of a change in capacity depends on how traffic redistributes across the network. Braess's paradox shows that even adding capacity can lengthen travel times \cite{ext:braess2005paradox}, and assigning more lanes to one direction of a road necessarily takes them from the other, so the traffic it sheds can burden the surrounding roads. This motivates deciding lane allocations with the whole network in view and, because speed limits shape how traffic uses the capacity that an allocation provides, deciding them together with the speed limits. The second difficulty is reachability. Assigning all lanes of a road to a single direction turns it into a one-way road, which can concentrate capacity where it is needed most but can also cut junctions off from each other. Robbins' theorem \cite{ext:robbins1939theorem} settles when every road of a network can be made one-way with all junctions still mutually reachable, but only for an orientation that never changes. We require a dynamic controller to preserve reachability at every time step, including while a reversal is in progress: before traffic may enter a lane in its new direction, the vehicles still traveling in its old direction must leave, and until they have, the lane serves neither direction \cite{ext:trb2004convertible}.

Prior work addresses parts of this problem. Dynamic lane reversal has been formulated as an optimization problem over a given or forecast demand, solved with integer and bi-level programs or within dynamic traffic assignment \cite{ext:hausknecht2011dlr,ext:duell2016dlr,ext:levin2016dlr,ext:chu2020dlr}. Reinforcement learning (RL) has been used to allocate lanes without a model of the demand, in experiments that keep at least one lane in each direction of every road \cite{ext:gunarathna2023clla}. Lane or road-width reallocation has been integrated with flow and speed control, but through model-based optimization on freeways and motorways \cite{ext:chen2022dlrtc,ext:jin2026ibc}. To our knowledge, no prior work learns a policy that jointly sets lane allocations and speed limits on every road of a general network while permitting one-way roads and guaranteeing that every junction remains reachable from every other.

We address this gap with a formulation that separates what a controller may propose from what takes effect. Every road has a \emph{road controller} that proposes, at every decision step, how many of its lanes serve each direction and a speed limit for each direction, bounded above by the nominal speed limit. All proposals pass through \emph{action resolution}, which accepts those that would close a direction one at a time, in random order, only as long as the network stays strongly connected, and otherwise relaxes them to keep one lane open or, on single-lane roads, reverts them. Together with a per-road restriction that keeps both directions of every bridge road open and an explicit model of lane clearing, this guarantees strong connectivity at every time step, including during the clearing that precedes a reversal. Like a shield in safe RL \cite{ext:alshiekh2018shielding}, action resolution thus lets the learned policy explore any admissible action without ever cutting a junction off from the rest of the network.

Because the traffic regulated by the controllers couples the whole network, we realize all road controllers with a single transformer \cite{ext:vaswani2017attention} whose tokens are road--time pairs. Its attention mechanism, \emph{Global Endpoint Relation-Aware Spatio-Temporal Self-Attention} (GERASTSA), attends jointly over space and time and, in the spirit of graph transformers \cite{ext:ying2021graphormer}, biases attention by the endpoint relations between roads. We pretrain the model on a self-supervised task native to the domain, forecasting the traffic of every road conditioned on the lane allocations and speed limits that take effect in between, and fine-tune it with proximal policy optimization (PPO) \cite{ext:schulman2017ppo} applied per road. Each road optimizes a \emph{mixed reward} that combines its own reward with those of its neighbors and of the whole network, so that the objective of a road also reflects the cost that its becoming one-way imposes on the roads that absorb its traffic.

Our contributions are as follows.

\begin{itemize}
\item We formulate network-wide dynamic lane allocation and directional speed limit control with one-way roads, explicit lane clearing, and a strong-connectivity requirement on the navigable graph (Section~\ref{sec:problem}).

\item We give an action-resolution algorithm, grounded in graph orientation theory, that enforces this requirement at every time step while leaving per-road admissibility to the action space, and show that it does so (Section~\ref{sec:problem:action-resolution}, Proposition~\ref{prop:connectivity}).

\item We propose the GERASTSA transformer, a single model for all road controllers whose attention is biased by the endpoint relations between roads, including the directions in which adjacent roads meet, and whose embeddings encode the junction degrees of every road and the position of every time step within the window (Section~\ref{sec:arch}).

\item We train it by pretraining on action-conditioned traffic forecasting and fine-tuning with per-road PPO under a mixed reward (Section~\ref{sec:exp}), and evaluate the learned policy on a grid network with commuter rush hours against a baseline without a policy and a heuristic that keeps every road two-way, against fine-tuning without pretraining, against restrictions of its actions at test time, and against ablations of the components of the GERASTSA transformer (Section~\ref{sec:results}).

\end{itemize}

The remainder of the paper is organized as follows. Section~\ref{sec:prior-work} reviews related work on lane reversal, VSLs, integrated and learning-based traffic control, and spatio-temporal and graph attention. Section~\ref{sec:problem} formulates the control problem and action resolution, Section~\ref{sec:arch} presents the GERASTSA transformer, and Section~\ref{sec:exp} describes the environment, pretraining, and fine-tuning. Section~\ref{sec:results} reports the results, Section~\ref{sec:discussion} discusses them, and Section~\ref{sec:conclusion} concludes.

%% file: sections/prior-work.tex
\section{Prior Work}\label{sec:prior-work}
We review work on lane reversal and VSLs, on their integration with other controls, on learning-based coordination of traffic controllers, and on attention models for spatio-temporal and graph-structured data, and then position our contribution.

\subsection{Lane Reversal and Contraflow}\label{sec:prior-work:lane-reversal}
Reversible lanes and contraflow operations match road capacity to directionally unbalanced demand, both for recurring commuter peaks and for emergency evacuation \cite{ext:trb2004convertible,ext:wolshon2001contraflow}. In practice, they follow fixed schedules or are activated for planned events \cite{ext:trb2004convertible}. At the network level, the choice of which roads to reverse has been formulated as an optimization problem, chiefly for evacuation, where the configuration is held for the duration of the event \cite{ext:kim2008contraflow}.

CAVs have motivated \emph{dynamic} lane reversal, in which lane directions follow the traffic as it evolves. \cite{ext:hausknecht2011dlr} compute lane configurations that maximize network flow with integer linear and bi-level programs, and examine how directional imbalance and the granularity of reallocation affect the benefit. \cite{ext:duell2016dlr} extend a system-optimal dynamic traffic assignment model based on the cell transmission model (CTM) so that it also determines lane directions at fine space--time resolution. \cite{ext:levin2016dlr} develop a CTM for dynamic lane reversal with autonomous vehicles and derive a heuristic from an analysis of a single link, which they apply to city networks. \cite{ext:chu2020dlr} optimize lane reversal together with the routes and schedules of individual vehicles as an integer program, solved with a distributed algorithm based on the alternating direction method of multipliers. For freeways, \cite{ext:frejo2016reversible} model reversible lanes macroscopically and operate them with a logic-based controller and with model predictive control (MPC). \cite{ext:li2024reversible} choose the locations, operating times, and installation modes of reversible lanes with a minimum-cost maximum-flow formulation that accounts for risk and economic costs.

These approaches optimize against a demand that is given or forecast in advance. \cite{ext:gunarathna2023clla} instead learn lane allocations with RL: local agents propose lane-direction changes, which are coordinated through a graph of dependencies between roads derived from vehicle paths, and connected vehicles are rerouted in response to the new configuration. In their experiments, every road keeps at least one lane in each direction. We likewise learn a policy that maps the observed state of the network to lane allocations, but we allow roads to become one-way whenever the network remains strongly connected, and we model the clearing of lanes explicitly, so that a reversal takes effect only once traffic allows.

Allowing roads to become one-way raises the question of reachability between junctions. Robbins' theorem \cite{ext:robbins1939theorem} characterizes the graphs that admit a strongly connected orientation; it extends to mixed multigraphs \cite{ext:boesch1980robbins}, and such orientations can be computed in linear time \cite{ext:chung1985strongly}. These results concern a single, static orientation. We use them to guarantee that an admissible initial configuration exists and combine them with action resolution so that strong connectivity is preserved at every time step while lane allocations change.

\subsection{Variable Speed Limits}\label{sec:prior-work:vsl}
VSLs lower the posted speed in response to traffic conditions to delay the onset of congestion, dampen shock waves, and improve safety \cite{ext:lu2014vsl}. \cite{ext:smulders1990vsl} found from field data that advisory variable speed signs barely changed capacity but stabilized traffic, and derived a control policy that postpones congestion on a freeway section. MPC coordinates speed limits with ramp metering through a macroscopic traffic model \cite{ext:hegyi2005vsl}, whereas the SPECIALIST algorithm derives speed-limit interventions from shock-wave theory to resolve traffic waves \cite{ext:hegyi2008specialist}; its field test on a Dutch freeway confirmed its effect on shock waves but also showed that it was less effective for other types of congestion \cite{ext:hegyi2010specialist}.

RL avoids the need for an explicit traffic model in the controller, although the policy is usually trained in simulation. \cite{ext:zhu2014vsl} formulate dynamic speed limit control on a network with stochastic demand and supply as a Markov decision process, and evaluate it on the Sioux Falls network in terms of throughput, delay, and emissions. On freeways, \cite{ext:li2017vsl} learn speed limits with Q-learning to relieve congestion at a recurrent bottleneck, \cite{ext:wu2020dvsl} learn lane-specific speed limits with a deep actor-critic algorithm in a setting with CAVs, and \cite{ext:yang2024dvsrg} represent the freeway as a graph of road topology, occupancy, and speed to assign speed limits per lane and location. We set speed limits on every road and direction of an urban network and decide them jointly with lane allocations.

\subsection{Integrated Control}\label{sec:prior-work:integrated}
Lane reversal has also been integrated with other traffic controls. \cite{ext:chen2022dlrtc} combine dynamic lane reversal with traffic-flow control on a motorway network: a multilane CTM and a mixed-integer linear program, solved within MPC, determine the lane directions and the control actions communicated to autonomous vehicles. For lane-free traffic of CAVs on bidirectional freeways, \cite{ext:jin2026ibc} integrate internal boundary control, which shifts the road width assigned to each direction, with ramp metering and VSLs, and minimize the total time spent with a quadratic program over an extended CTM. These studies indicate that, on freeways and motorways, reallocating directional capacity and regulating flow can complement each other. They rely on optimization over a macroscopic traffic model and consider freeway or motorway settings with continuous width or bidirectional roads, whereas we consider a general road network with integer lane counts, one-way roads, and an explicit requirement on directed reachability.

\subsection{Learning-Based Network Traffic Control}\label{sec:prior-work:rl}
Network-wide traffic control with RL has been studied most extensively for traffic signals. Every intersection typically acts as an agent, and the agents must cooperate because their decisions affect each other's traffic. \cite{ext:chu2020marl} let agents share information with their neighbors and discount the rewards of distant agents, \cite{ext:wei2019colight} let each intersection aggregate the states of its neighbors with graph attention, and \cite{ext:devailly2022igrl} share a single graph-convolutional policy across all intersections, which transfers without retraining to road networks and traffic conditions not seen in training, including the 3,971 signals of Manhattan. Like the last, our model shares its parameters across all roads and is not tied to the size of the network (Section~\ref{sec:arch}), and our mixed reward adapts the reward discounting of \cite{ext:chu2020marl} (Section~\ref{sec:exp:fine-tuning:reward}).

More broadly, independent and multi-agent variants of PPO, in which agents with shared parameters optimize their own clipped objectives, perform strongly on cooperative benchmarks \cite{ext:dewitt2020ippo,ext:yu2022mappo}. We adopt this per-agent treatment, but whereas the policy of each agent acts on a local observation in these methods, ours observes the whole network through a single transformer, as does our critic. Invalid action masking removes inadmissible discrete actions from the policy \cite{ext:huang2022masking}, and shielding replaces unsafe actions with safe ones before they reach the environment \cite{ext:alshiekh2018shielding}; we use the former for per-road admissibility and the latter, in the form of action resolution, for the network-wide connectivity requirement.

Finally, self-supervised objectives can make RL more data-efficient, either as auxiliary tasks trained alongside the policy \cite{ext:jaderberg2017unreal} or as a pretraining stage that learns representations, including models of the environment dynamics, before fine-tuning \cite{ext:schwarzer2021pretraining}. Our pretraining follows the latter approach with action-conditioned traffic forecasting, a task native to the traffic domain (Section~\ref{sec:exp:pretraining}).

\subsection{Spatio-Temporal and Graph Attention}\label{sec:prior-work:attention}
Traffic forecasting models capture spatial dependencies on the road network together with temporal dynamics, originally by combining graph convolutions with recurrent or convolutional sequence models \cite{ext:li2018dcrnn,ext:yu2018stgcn}. Attention-based models have since become prominent. GMAN stacks blocks of spatial and temporal attention in an encoder--decoder architecture \cite{ext:zheng2020gman}, and PDFormer applies spatial self-attention with graph masks for short- and long-range dependencies and accounts for the delay with which traffic propagates between locations \cite{ext:jiang2023pdformer}. Unlike these models, which attend over space and time in separate steps, our model attends jointly over all road--time pairs of a window (Section~\ref{sec:arch}).

The transformer \cite{ext:vaswani2017attention} is permutation-equivariant and therefore needs structural information to be injected. Relative position representations add learned terms for clipped relative distances to the attention computation \cite{ext:shaw2018relative}, and Graphormer adds a learnable bias indexed by the shortest-path distance between two nodes to the attention scores and a degree-based centrality encoding to the node features \cite{ext:ying2021graphormer}. Our attention follows this design on the controller graph but refines the relations between adjacent roads by the directions in which both flow into their shared junction, and our spatial embedding refines the degree encoding by direction, since both distinctions bear on how traffic passes from one road to the next. The choice of graph matters: intersection adjacency \cite{ext:wei2019colight}, shared vehicle paths \cite{ext:gunarathna2023clla}, and adjacency between controlled roads encode different relationships. We use the last, which matches our decision variables, and keep it separate from the navigable graph, which describes the movements that remain physically available.

\subsection{Positioning}\label{sec:prior-work:positioning}
In summary, prior work has controlled lane directions dynamically, learned lane allocations with RL, and integrated lane or road-width reallocation with flow and speed control through model-based optimization. As noted in Section~\ref{sec:intro}, however, to our knowledge none of it combines these elements in a learned policy for general networks with one-way roads and guaranteed strong connectivity. Our approach does so through action resolution grounded in graph orientation theory and a single Graphormer-style transformer for all road controllers.

%% file: sections/problem-formulation.tex
\section{Problem Formulation}\label{sec:problem}

\begin{figure*}[t]
\centering
\begin{minipage}[b]{\dimexpr(\textwidth-24pt)/3\relax}\centering\includegraphics{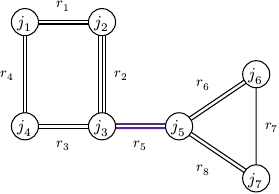}\end{minipage}\hspace{12pt}\begin{minipage}[b]{\dimexpr(\textwidth-24pt)/3\relax}\centering\includegraphics{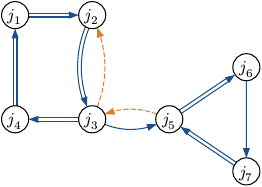}\end{minipage}\hspace{12pt}\begin{minipage}[b]{\dimexpr(\textwidth-24pt)/3\relax}\centering\includegraphics{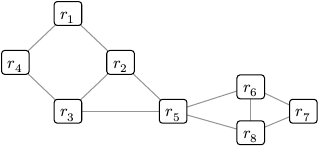}\end{minipage}\\[12pt]
\begin{minipage}[t]{\dimexpr(\textwidth-24pt)/3\relax}\centering\footnotesize (a) Infrastructure graph $G$\end{minipage}\hspace{12pt}\begin{minipage}[t]{\dimexpr(\textwidth-24pt)/3\relax}\centering\footnotesize (b) Navigable graph $\vec{G}^{(t)}$\end{minipage}\hspace{12pt}\begin{minipage}[t]{\dimexpr(\textwidth-24pt)/3\relax}\centering\footnotesize (c) Controller graph $L (G)$\end{minipage}
\caption{An example network with $|\mathcal{J}| = 7$ junctions and $|\mathcal{R}| = 8$ roads. (a) Each stroke is one lane. Road $r_{5}$ (heavy) is the only bridge road, $\mathcal{R}_{\leftrightarrow} = \{r_{5}\}$, so it has at least two lanes, and strong connectivity requires it to keep at least one open lane in each direction. (b) A strongly connected navigable arc set $\vec{\mathcal{R}}^{(t)} \in \mathcal{S}$: arrows show the navigable arcs (solid: forward; dashed: backward), with one stroke per open lane $\bar{\ell}_{r, d}^{(t)}$; most roads are one-way, yet $\vec{G}^{(t)}$ is strongly connected. (c) One road controller per road; controllers of roads that share a junction are neighbors.}
\label{fig:graphs}
\end{figure*}

\subsection{Road Graphs}\label{sec:problem:road-graphs}
Robbins' theorem \cite{ext:robbins1939theorem} states that a connected undirected graph admits a strongly connected orientation, that is, a direction for every edge such that every vertex can reach every other vertex, if and only if the graph has no bridge, an edge whose removal disconnects it. Operating every road as a one-way road is such an orientation, so, with reachability judged between junctions, the theorem characterizes exactly when a road network can run entirely on one-way roads with every junction still reachable from every other. Roads, however, consist of lanes, and a road with two or more lanes can serve both directions at once. We build on this observation: the lanes of each road may be split between its two directions, and roads that would otherwise violate the conditions of the theorem are required to have enough lanes for such a split.

We model the road infrastructure as the \emph{infrastructure graph} $G$, a simple undirected graph whose vertices are a finite set of \emph{junctions} $\mathcal{J}$ and whose edges are \emph{roads} $\mathcal{R}$, each road being an unordered pair of distinct junctions. The network has at least one road, and each road $r \in \mathcal{R}$ has $L_{r} \in \mathbb{Z}_{\geq 1}$ \emph{total lanes}. Fig.~\ref{fig:graphs} (a) shows an example.
\begin{equation}\label{eq:infrastructure-graph}
\begin{aligned}
& G = (\mathcal{J}, \mathcal{R}); \quad \mathcal{R} \subseteq \binom{\mathcal{J}}{2}; \quad |\mathcal{R}| \geq 1;
\end{aligned}
\end{equation}

A \emph{cut} is a nonempty proper subset of junctions, and $\mathcal{C} = 2^{\mathcal{J}} \setminus \{\varnothing, \mathcal{J}\}$ is the set of all cuts. The \emph{cut-set} $\mathcal{B} (\mathcal{X})$ of a cut $\mathcal{X}$ consists of the roads with exactly one endpoint in $\mathcal{X}$.
\begin{equation}\label{eq:cut-set}
\begin{aligned}
& \mathcal{B} : 2^{\mathcal{J}} \to 2^{\mathcal{R}}; \quad \mathcal{B} (\mathcal{X}) = \mleft\{r \in \mathcal{R} \mathrel{}\middle|\mathrel{} |r \cap \mathcal{X}| = 1\mright\};
\end{aligned}
\end{equation}
Requiring every cut-set to be nonempty is equivalent to requiring $G$ to be connected.
\begin{equation}\label{eq:connected}
\begin{aligned}
& \mathcal{B} (\mathcal{X}) \neq \varnothing\ \forall \mathcal{X} \in \mathcal{C};
\end{aligned}
\end{equation}
A \emph{bridge road} is a road that forms the entire cut-set of some cut, and $\mathcal{R}_{\leftrightarrow}$ is the set of bridge roads.
\begin{equation}\label{eq:bridge-road-set}
\begin{aligned}
& \mathcal{R}_{\leftrightarrow} = \mleft\{r \in \mathcal{R} \mathrel{}\middle|\mathrel{} \exists \mathcal{X} \in \mathcal{C} : \mathcal{B} (\mathcal{X}) = \{r\}\mright\};
\end{aligned}
\end{equation}
In line with Robbins' theorem, we assume that every bridge road has at least two lanes so that it can carry traffic in both directions at once, which a single-lane road cannot do.
\begin{equation}\label{eq:bridge-roads}
\begin{aligned}
& L_{r} \geq 2\ \forall r \in \mathcal{R}_{\leftrightarrow};
\end{aligned}
\end{equation}

Traffic on a road flows in one of two \emph{directions}, $\mathcal{D} = \{\mathrm{fwd}, \mathrm{bwd}\}$. We fix a reference ordering of the endpoints of every road, marked by the superscripts $-$ and $+$: traveling from its first endpoint $j_{r}^{-}$ to its second endpoint $j_{r}^{+}$ is \emph{forward}, and the opposite is \emph{backward}. The \emph{orientation map} $\omega$ sends a road and a direction to the corresponding \emph{arc}, an ordered pair of junctions. Throughout, an arc is an oriented road, and $\omega$ applied to a set denotes the set of images.
\begin{equation}\label{eq:orientation}
\begin{aligned}
& r = \{j_{r}^{-}, j_{r}^{+}\}\ \mathrm{with\ fixed\ ordering}\ (j_{r}^{-}, j_{r}^{+})\ \forall r \in \mathcal{R}; \\
& \omega : \mathcal{R} \times \mathcal{D} \to \mathcal{J}^{2}; \quad \omega (r, d) = \begin{cases}
(j_{r}^{-}, j_{r}^{+}) & \mathrm{if}\ d = \mathrm{fwd} \\
(j_{r}^{+}, j_{r}^{-}) & \mathrm{if}\ d = \mathrm{bwd}
\end{cases};
\end{aligned}
\end{equation}

We consider a horizon of $T \in \mathbb{Z}_{\geq 1}$ steps, with \emph{time steps} $\breve{\mathcal{T}} = \{0, \dots, T\}$ and \emph{decision steps} $\mathcal{T} = \breve{\mathcal{T}} \setminus \{T\}$, the time steps at which the controllers act. A superscript $(t)$ marks the value of a quantity at time step $t$. At time step $t \in \breve{\mathcal{T}}$, road $r \in \mathcal{R}$ has $\bar{\ell}_{r, d}^{(t)} \in \{0, \dots, L_{r}\}$ \emph{open lanes} usable in direction $d \in \mathcal{D}$, defined in \eqref{eq:open-lanes}. The arcs with at least one open lane form the \emph{navigable arc set} $\vec{\mathcal{R}}^{(t)}$, which defines the \emph{navigable graph} $\vec{G}^{(t)}$ on the junctions (Fig.~\ref{fig:graphs} (b)).
\begin{equation}\label{eq:navigable}
\begin{aligned}
& \vec{\mathcal{R}}^{(t)} = \omega \mleft(\mleft\{(r, d) \in \mathcal{R} \times \mathcal{D} \mathrel{}\middle|\mathrel{} \bar{\ell}_{r, d}^{(t)} > 0\mright\}\mright), \\
& \vec{G}^{(t)} = \mleft(\mathcal{J}, \vec{\mathcal{R}}^{(t)}\mright)\ \forall t \in \breve{\mathcal{T}};
\end{aligned}
\end{equation}

We require that drivers be able to travel between any two junctions at every time step, that is, that the navigable graph be strongly connected. An arc set has this property exactly when, for every cut $\mathcal{X}$, it contains an arc from $\mathcal{X}$ to its complement $\mathcal{X}^{c} = \mathcal{J} \setminus \mathcal{X}$. Because the complement of a cut is again a cut, the same condition also guarantees an arc into every cut. We collect all \emph{strongly connected arc sets} in $\mathcal{S}$ and require the navigable arc set to belong to $\mathcal{S}$ at every time step.
\begin{equation}\label{eq:strong-connectivity}
\begin{aligned}
& \mathcal{S} = \\
&\ \mleft\{\mathcal{Y} \subseteq \omega (\mathcal{R} \times \mathcal{D}) \mathrel{}\middle|\mathrel{} \forall \mathcal{X} \in \mathcal{C} : \mathcal{Y} \cap (\mathcal{X} \times \mathcal{X}^{c}) \neq \varnothing\mright\}; \\
& \vec{\mathcal{R}}^{(t)} \in \mathcal{S}\ \forall t \in \breve{\mathcal{T}};
\end{aligned}
\end{equation}

At $t = 0$, this \emph{strong-connectivity requirement} is an assumption on the initial configuration; at later time steps, it must be maintained by action resolution (Section~\ref{sec:problem:action-resolution}). The assumptions on $G$ in \eqref{eq:connected} and \eqref{eq:bridge-roads} ensure that the requirement can be met. By Robbins' theorem, the roads within each connected component of $G$ without its bridge roads, which has no bridges, admit a strongly connected orientation, and opening both directions of every bridge road, which is possible because it has at least two lanes, joins these components into a strongly connected whole. Equivalently, treating every bridge road as a pair of opposite arcs and every other road as an undirected edge yields a mixed graph in which no undirected edge is a bridge, which admits a strongly connected orientation by the extension of Robbins' theorem to mixed multigraphs \cite{ext:boesch1980robbins}; such an orientation, and hence an admissible initial configuration, can be found in linear time \cite{ext:chung1985strongly}. Moreover, $\mathcal{S}$ is closed under supersets: opening additional arcs never breaks strong connectivity, a fact we rely on during action resolution.

\subsection{Controller Graph}\label{sec:problem:controller-graph}
Every road is governed by a \emph{road controller} that decides the road's lane allocation and speed limits. Two roads that share a junction exchange traffic directly, so their controllers are \emph{neighbors}. The resulting coupling structure is the \emph{controller graph}, the line graph $L (G)$ of the infrastructure graph: its vertices are the roads, and its edges $\mathcal{E}$ join pairs of roads with a common endpoint (Fig.~\ref{fig:graphs} (c)).
\begin{equation}\label{eq:controller-graph}
\begin{aligned}
& \mathcal{E} = \mleft\{\{r_{1}, r_{2}\} \in \binom{\mathcal{R}}{2} \mathrel{}\middle|\mathrel{} r_{1} \cap r_{2} \neq \varnothing\mright\}; \\
& L (G) = (\mathcal{R}, \mathcal{E});
\end{aligned}
\end{equation}

The controller graph describes how roads are coupled directly through traffic, not which controllers may share information; in our model (Section~\ref{sec:arch}), every controller draws on the state of the whole network. The strong-connectivity requirement in \eqref{eq:strong-connectivity}, by contrast, couples all roads at once and cannot be checked within any single neighborhood, so it is enforced during action resolution (Section~\ref{sec:problem:action-resolution}).

\subsection{Lane Allocations and Speed Limits}\label{sec:problem:lanes-speeds}
\begin{figure}[tb]
\centering
\includegraphics{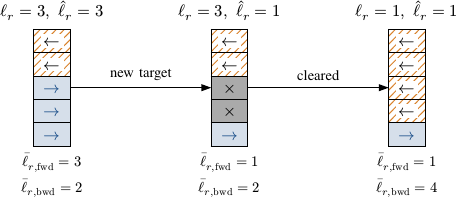}
\caption{Clearing two lanes of a road with $L_{r} = 5$ (lane 1, the first forward lane, at the bottom). When the target lanes $\hat{\ell}_{r}$ drop below the active lanes $\ell_{r}$, lanes $\hat{\ell}_{r} + 1, \dots, \ell_{r}$ are closed to both directions (dark grey) while vehicles traveling in their old direction leave them. Clearing is atomic: once it completes, all of these lanes open at once to the backward direction (hatched; forward lanes are tinted) and $\ell_{r} = \hat{\ell}_{r}$. The open lanes $\bar{\ell}_{r, d}$ below each state need not sum to $L_{r}$ while the road is clearing.}
\label{fig:lane-allocation}
\end{figure}

The lanes of a road are allocated between its two directions, and we describe an allocation by the number of lanes assigned to the forward direction; the remaining lanes are assigned to the backward direction, as in reversible-lane and contraflow operations \cite{ext:trb2004convertible} and dynamic lane reversal \cite{ext:levin2016dlr}. At time step $t$, road $r$ has $\ell_{r}^{(t)}$ \emph{active lanes} and $\hat{\ell}_{r}^{(t)}$ \emph{target lanes}, both counted in the forward direction, and every road starts at its target, $\hat{\ell}_{r}^{(0)} = \ell_{r}^{(0)}$. Whenever the two differ, the lanes in between must be \emph{cleared} before they can change direction: they are closed in both directions until the vehicles traveling in their old direction have left, so that vehicles in the new direction can safely enter, as in the operation of reversible lanes in practice \cite{ext:trb2004convertible}. The open lanes in a direction are therefore those assigned to it by both the active and the target lanes, and the forward and backward open lanes need not sum to $L_{r}$. The roads with lanes being cleared are the \emph{clearing roads} $\mathcal{R}_{\approx}^{(t)}$. Clearing is atomic: once it completes, all clearing lanes of the road open at once in their new direction, and the active lanes become equal to the target lanes. Fig.~\ref{fig:lane-allocation} illustrates this process.
\begin{equation}\label{eq:open-lanes}
\begin{aligned}
& \ell_{r}^{(t)}, \hat{\ell}_{r}^{(t)} \in \{0, \dots, L_{r}\}\ \forall r \in \mathcal{R}\ \forall t \in \breve{\mathcal{T}}; \\
& \bar{\ell}_{r, d}^{(t)} = \begin{cases}
\min \mleft(\ell_{r}^{(t)}, \hat{\ell}_{r}^{(t)}\mright) & \mathrm{if}\ d = \mathrm{fwd} \\
L_{r} - \max \mleft(\ell_{r}^{(t)}, \hat{\ell}_{r}^{(t)}\mright) & \mathrm{if}\ d = \mathrm{bwd}
\end{cases} \\
&\ \forall (r, d) \in \mathcal{R} \times \mathcal{D}\ \forall t \in \breve{\mathcal{T}}; \\
& \mathcal{R}_{\approx}^{(t)} = \mleft\{r \in \mathcal{R} \mathrel{}\middle|\mathrel{} \ell_{r}^{(t)} \neq \hat{\ell}_{r}^{(t)}\mright\}\ \forall t \in \breve{\mathcal{T}};
\end{aligned}
\end{equation}

Each road controller also sets a speed limit for each direction of its road. The \emph{nominal speed limit} $s_{r, d}^{(0)}$ is the speed limit set by traffic engineers based on safety, road geometry, zoning, and similar considerations. It is an upper bound: the controller may lower the speed limit to regulate traffic, as in VSL control \cite{ext:hegyi2005vsl,ext:lu2014vsl}, but never raise it beyond the nominal value, so the posted speed never exceeds the value that traffic engineers set for the road. From below, speed limits are bounded by a common \emph{crawl speed} $s_{\min} \in \mathbb{R}_{> 0}$, which every nominal speed limit exceeds. We parametrize the \emph{active speed limit} $s_{r, d}^{(t)}$ by the \emph{active normalized speed limit} $u_{r, d}^{(t)} \in [0, 1]$, which interpolates linearly between the crawl speed at $0$ and the nominal speed limit at $1$; initially, every active speed limit is nominal.
\begin{equation}\label{eq:speed-limits}
\begin{aligned}
& u_{r, d}^{(0)} = 1,\ s_{r, d}^{(0)} \in \mathbb{R}_{> s_{\min}}\ \forall (r, d) \in \mathcal{R} \times \mathcal{D}; \\
& u_{r, d}^{(t)} \in [0, 1],\ s_{r, d}^{(t)} = s_{\min} + u_{r, d}^{(t)} \mleft(s_{r, d}^{(0)} - s_{\min}\mright) \\
&\ \forall (r, d) \in \mathcal{R} \times \mathcal{D}\ \forall t \in \breve{\mathcal{T}};
\end{aligned}
\end{equation}

Since the crawl speed is strictly positive, lowering a speed limit slows the traffic in a direction but never stops it; traffic in a direction can only be stopped by closing all of its lanes. Vehicles on clearing lanes still travel in their old direction and remain subject to its active speed limit until they leave.

\subsection{Joint Action}\label{sec:problem:joint-action}
At every decision step, the road controllers jointly select an \emph{action} consisting of \emph{proposed lanes} $\tilde{\ell}^{(t)}$, one forward lane count per road, and \emph{proposed normalized speed limits} $\tilde{u}^{(t)}$, one per road and direction. The action space is the product of the \emph{admissible proposals} for lanes and speed limits.
\begin{equation}\label{eq:action}
\begin{aligned}
& \mathcal{A}^{(t)} = \tilde{\mathcal{L}}^{(t)} \times \tilde{\mathcal{U}}^{(t)}, \\
& a^{(t)} = \mleft(\tilde{\ell}^{(t)}, \tilde{u}^{(t)}\mright) \in \mathcal{A}^{(t)}\ \forall t \in \mathcal{T};
\end{aligned}
\end{equation}

Admissible proposed lanes are defined per road. A clearing road is locked to its target: its only admissible proposal is its target lanes, so a road cannot be retargeted until its clearing completes. A bridge road that is not clearing must keep at least one lane in each direction, which excludes the extreme proposals $0$ and $L_{r}$. Every other road may propose any allocation, including one that assigns all of its lanes to a single direction.
\begin{equation}\label{eq:admissible-lanes}
\begin{aligned}
& \tilde{\mathcal{L}}_{r}^{(t)} = \begin{cases}
\mleft\{\hat{\ell}_{r}^{(t)}\mright\} & \mathrm{if}\ r \in \mathcal{R}_{\approx}^{(t)} \\
\{1, \dots, L_{r} - 1\} & \mathrm{if}\ r \in \mathcal{R}_{\leftrightarrow} \setminus \mathcal{R}_{\approx}^{(t)} \\
\{0, \dots, L_{r}\} & \mathrm{otherwise}
\end{cases} \\
&\ \forall r \in \mathcal{R}\ \forall t \in \mathcal{T}; \\
& \tilde{\mathcal{L}}^{(t)} = \bigtimes_{r \in \mathcal{R}} \tilde{\mathcal{L}}_{r}^{(t)}\ \forall t \in \mathcal{T};
\end{aligned}
\end{equation}

Admissible proposed normalized speed limits are defined per road and direction. A direction with open lanes may receive any proposed normalized speed limit in $[0, 1]$, whereas the proposed normalized speed limit of a direction without open lanes is frozen at its active normalized speed limit, since no vehicle can enter that direction.
\begin{equation}\label{eq:admissible-speeds}
\begin{aligned}
& \tilde{\mathcal{U}}_{r, d}^{(t)} = \begin{cases}
\mleft\{u_{r, d}^{(t)}\mright\} & \mathrm{if}\ \bar{\ell}_{r, d}^{(t)} = 0 \\
[0, 1] & \mathrm{otherwise}
\end{cases} \\
&\ \forall (r, d) \in \mathcal{R} \times \mathcal{D}\ \forall t \in \mathcal{T}; \\
& \tilde{\mathcal{U}}^{(t)} = \bigtimes_{(r, d) \in \mathcal{R} \times \mathcal{D}} \tilde{\mathcal{U}}_{r, d}^{(t)}\ \forall t \in \mathcal{T};
\end{aligned}
\end{equation}

Every admissible set in \eqref{eq:admissible-lanes} and \eqref{eq:admissible-speeds} is nonempty; for bridge roads that are not clearing, this relies on $L_{r} \geq 2$ from \eqref{eq:bridge-roads}. Hence $\mathcal{A}^{(t)} \neq \varnothing$ for all $t \in \mathcal{T}$.

Admissibility is a per-road property and does not depend on the proposals of other roads. In particular, the action space encodes the strong-connectivity requirement in \eqref{eq:strong-connectivity}, which couples all roads, only through the bridge-road restriction. A closing proposal of a bridge road would always be rejected, so excluding it from the action space leaves less to action resolution and makes the effect of an action more predictable. All other actions that would violate the requirement are filtered during action resolution.

\subsection{Action Resolution}\label{sec:problem:action-resolution}
Proposed lanes do not take effect immediately: an accepted proposal becomes the road's next target lanes, and the road then clears toward them. A new target closes the lanes on which it disagrees with the active lanes at once, but these lanes reopen in their new direction only once clearing completes, at a time determined by traffic. We therefore assess proposals in the state before clearing completes, which has the fewest open lanes; completing clearing can only add to them. To this end, we consider a profile $n$ of target lanes, with $n_{r}$ for road $r$. The \emph{prospective open lanes} $\mu_{r, d}^{(t)} (n_{r})$ are the open lanes of road $r$ in direction $d$ if its target lanes were $n_{r}$ while its current active lanes $\ell_{r}^{(t)}$ remained;
\begin{equation}\label{eq:prospective}
\begin{aligned}
& \mu_{r, d}^{(t)} : \{0, \dots, L_{r}\} \to \{0, \dots, L_{r}\}, \\
& \mu_{r, d}^{(t)} (n_{r}) = \begin{cases}
\min \mleft(\ell_{r}^{(t)}, n_{r}\mright) & \mathrm{if}\ d = \mathrm{fwd} \\
L_{r} - \max \mleft(\ell_{r}^{(t)}, n_{r}\mright) & \mathrm{if}\ d = \mathrm{bwd}
\end{cases} \\
&\ \forall (r, d) \in \mathcal{R} \times \mathcal{D}\ \forall t \in \breve{\mathcal{T}};
\end{aligned}
\end{equation}
in particular, by \eqref{eq:open-lanes}, $\mu_{r, d}^{(t)} \mleft(\hat{\ell}_{r}^{(t)}\mright) = \bar{\ell}_{r, d}^{(t)}$. The \emph{prospective navigable arc set} $\mathcal{P}^{(t)} (n)$ contains the arcs that would have at least one open lane.
\begin{equation}\label{eq:prospective-arcs}
\begin{aligned}
& \mathcal{P}^{(t)} : \bigtimes_{r \in \mathcal{R}} \{0, \dots, L_{r}\} \to 2^{\omega (\mathcal{R} \times \mathcal{D})}, \\
& \mathcal{P}^{(t)} (n) = \omega \mleft(\mleft\{(r, d) \in \mathcal{R} \times \mathcal{D} \mathrel{}\middle|\mathrel{} \mu_{r, d}^{(t)} (n_{r}) > 0\mright\}\mright) \\
&\ \forall t \in \breve{\mathcal{T}};
\end{aligned}
\end{equation}
Finally, for a road with at least two lanes, the \emph{relaxation function} $\rho_{r}$ maps each extreme allocation, $0$ or $L_{r}$, to the nearest allocation that leaves one lane in the opposite direction.
\begin{equation}\label{eq:relaxation}
\begin{aligned}
& \rho_{r} : \{0, L_{r}\} \to \{1, L_{r} - 1\}, \\
& \rho_{r} (n_{r}) = \begin{cases}
1 & \mathrm{if}\ n_{r} = 0 \\
L_{r} - 1 & \mathrm{if}\ n_{r} = L_{r}
\end{cases} \\
&\ \forall r \in \mathcal{R}\ \mathrm{with}\ L_{r} \geq 2;
\end{aligned}
\end{equation}

We call a road's proposal \emph{closing} if it would close a direction that is currently open, that is, if $\bar{\ell}_{r, d}^{(t)} > 0$ but $\mu_{r, d}^{(t)} \mleft(\tilde{\ell}_{r}^{(t)}\mright) = 0$ for some direction $d$. An open forward direction implies $\ell_{r}^{(t)} \geq 1$, so it closes only for the proposal $0$; likewise, an open backward direction implies $\ell_{r}^{(t)} \leq L_{r} - 1$, so it closes only for the proposal $L_{r}$. Hence every closing proposal is extreme, $\tilde{\ell}_{r}^{(t)} \in \{0, L_{r}\}$. By \eqref{eq:admissible-lanes}, neither a clearing road nor a bridge road ever proposes closing: the former proposes its target lanes, whose prospective open lanes are its open lanes, and the latter cannot propose $0$ or $L_{r}$. A non-closing proposal keeps every open direction open and therefore never removes an arc from the prospective navigable arc set. Algorithm~\ref{alg:rejection} exploits this: it accepts all non-closing proposals at once, then considers the closing roads one at a time in uniformly random order, tentatively accepting each proposal and rejecting it if the prospective navigable arc set would leave $\mathcal{S}$. A rejected proposal is relaxed rather than discarded: a road with at least two lanes adopts the \emph{relaxed proposal} $\rho_{r} \mleft(\tilde{\ell}_{r}^{(t)}\mright)$, which keeps one lane in the direction the proposal would have closed, whereas a single-lane road reverts to its current target lanes. The algorithm builds the next target lanes $\hat{\ell}^{(t + 1)}$ in place, starting from the current target lanes $\hat{\ell}^{(t)}$. The random order prevents any road from being systematically favored when closing proposals conflict, and it makes the transition from one time step to the next stochastic, together with traffic. Algorithm~\ref{alg:rejection} is a shield \cite{ext:alshiekh2018shielding} in the sense of Section~\ref{sec:prior-work:rl}.

Closing proposals conflict when each keeps the network strongly connected on its own but not together with the others, which is why they are checked one at a time against the proposals already accepted. Suppose, unlike in Fig.~\ref{fig:graphs} (b), that every road of the left cycle of the example network has both directions open and is not clearing, and that $r_{1}$ proposes $L_{r_{1}} = 2$, all of its lanes forward from $j_{1}$ to $j_{2}$, and $r_{3}$ proposes $0$, all of its lanes backward from $j_{4}$ to $j_{3}$, while all other roads propose their target lanes. Either closing proposal alone leaves the network strongly connected, but together they leave no arc from the cut $\{j_{2}, j_{3}, j_{5}, j_{6}, j_{7}\}$ to its complement. Whichever of the two comes first in the random order is accepted, and the other is relaxed to one lane in each direction.

\begin{algorithm}[tb]
\caption{Resolution of lane proposals at decision step $t \in \mathcal{T}$.}\label{alg:rejection}
\begin{algorithmic}[1]
\Statex \textbf{Input:} active lanes $\ell^{(t)}$, target lanes $\hat{\ell}^{(t)}$, proposed lanes $\tilde{\ell}^{(t)}$
\Statex \textbf{Output:} next target lanes $\hat{\ell}^{(t + 1)}$
\State start from the target lanes: $\hat{\ell}^{(t + 1)} \leftarrow \hat{\ell}^{(t)}$
\State start with an empty list of closing roads
\Block{\textbf{for each} road $r \in \mathcal{R}$ \textbf{do}}
\Block{\textbf{if} $\bar{\ell}_{r, d}^{(t)} > 0 \Rightarrow \mu_{r, d}^{(t)} \mleft(\tilde{\ell}_{r}^{(t)}\mright) > 0$ \textbf{for all} $d \in \mathcal{D}$ \textbf{then}}
\State accept the non-closing proposal: $\hat{\ell}_{r}^{(t + 1)} \leftarrow \tilde{\ell}_{r}^{(t)}$
\EndBlock
\Block{\textbf{else}}
\State append $r$ to the list of closing roads
\EndBlock
\EndBlock
\State shuffle the closing roads uniformly at random
\Block{\textbf{for each} closing road $r$, in shuffled order, \textbf{do}}
\State tentatively accept the proposal: $\hat{\ell}_{r}^{(t + 1)} \leftarrow \tilde{\ell}_{r}^{(t)}$
\Block{\textbf{if} $\mathcal{P}^{(t)} \mleft(\hat{\ell}^{(t + 1)}\mright) \notin \mathcal{S}$ \textbf{then}}
\Block{\textbf{if} $L_{r} \geq 2$ \textbf{then}}
\State relax the proposal: $\hat{\ell}_{r}^{(t + 1)} \leftarrow \rho_{r} \mleft(\tilde{\ell}_{r}^{(t)}\mright)$
\EndBlock
\Block{\textbf{else}}
\State revert to the target lanes: $\hat{\ell}_{r}^{(t + 1)} \leftarrow \hat{\ell}_{r}^{(t)}$
\EndBlock
\EndBlock
\EndBlock
\State \textbf{return} $\hat{\ell}^{(t + 1)}$
\end{algorithmic}
\end{algorithm}

\begin{figure}[tb]
\centering
\includegraphics{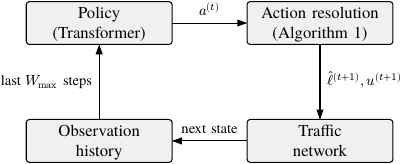}
\caption{One decision step. The policy of the road controllers proposes an action $a^{(t)}$, consisting of proposed lanes and proposed normalized speed limits for every road. Action resolution rejects closing proposals that would break strong connectivity, relaxing them on multi-lane roads and reverting them on single-lane roads, yielding the next target lanes $\hat{\ell}^{(t + 1)}$; proposed normalized speed limits are always accepted and become the next active normalized speed limits $u^{(t + 1)}$. The traffic network applies the resulting active speed limits immediately, and each clearing road completes clearing once its clearing lanes are empty.}
\label{fig:control-loop}
\end{figure}

Proposed normalized speed limits are always accepted and become the next active normalized speed limits. Each road then either keeps its current active lanes or completes clearing, in which case its next active lanes are its next target lanes; which of the two occurs is determined by traffic, not by the controllers (Fig.~\ref{fig:control-loop}).
\begin{equation}\label{eq:transition}
\begin{aligned}
& u_{r, d}^{(t + 1)} = \tilde{u}_{r, d}^{(t)}\ \forall (r, d) \in \mathcal{R} \times \mathcal{D}\ \forall t \in \mathcal{T}; \\
& \ell_{r}^{(t + 1)} \in \mleft\{\ell_{r}^{(t)}, \hat{\ell}_{r}^{(t + 1)}\mright\}\ \forall r \in \mathcal{R}\ \forall t \in \mathcal{T};
\end{aligned}
\end{equation}

\begin{proposition}\label{prop:connectivity}
If $\vec{\mathcal{R}}^{(0)} \in \mathcal{S}$, then $\vec{\mathcal{R}}^{(t)} \in \mathcal{S}$ for all $t \in \breve{\mathcal{T}}$, and the navigable arc set stays in $\mathcal{S}$ between consecutive time steps, for any admissible actions $a^{(t)} \in \mathcal{A}^{(t)}$, any order of the closing roads in Algorithm~\ref{alg:rejection}, and any times at which clearing completes.
\end{proposition}

\emph{Proof:} We show that the requirement carries over from time step $t \in \mathcal{T}$, where $\vec{\mathcal{R}}^{(t)} \in \mathcal{S}$, to time step $t + 1$ and to the period in between, in two stages.

First, the prospective navigable arc set $\mathcal{P}^{(t)} \mleft(\hat{\ell}^{(t + 1)}\mright)$ remains in $\mathcal{S}$ throughout Algorithm~\ref{alg:rejection}. Initially, it equals $\vec{\mathcal{R}}^{(t)} \in \mathcal{S}$, because $\mu_{r, d}^{(t)} \mleft(\hat{\ell}_{r}^{(t)}\mright) = \bar{\ell}_{r, d}^{(t)}$ for every road and direction. Accepting a non-closing proposal leaves the arc set unchanged: it keeps every open direction open, and no proposal opens a closed direction before clearing completes. A closing proposal is kept only if the arc set stays in $\mathcal{S}$. When a closing proposal is rejected, reverting restores the road's open lanes, and therefore the preceding arc set in $\mathcal{S}$. Relaxing restores the preceding arc set as well, because the relaxed proposal is neither $0$ nor $L_{r}$ and is therefore not closing. Hence, once the algorithm returns, $\mathcal{P}^{(t)} \mleft(\hat{\ell}^{(t + 1)}\mright) \in \mathcal{S}$.

Second, the next navigable arc set $\vec{\mathcal{R}}^{(t + 1)}$ contains $\mathcal{P}^{(t)} \mleft(\hat{\ell}^{(t + 1)}\mright)$: by \eqref{eq:transition}, a road that does not complete clearing has exactly its prospective open lanes at the next time step, whereas a road that completes clearing opens all of its clearing lanes, which only adds open lanes. Since $\mathcal{S}$ is closed under supersets, $\vec{\mathcal{R}}^{(t + 1)} \in \mathcal{S}$. The same holds between the two time steps: the next target lanes and next active speed limits take effect immediately, so the open lanes in that period are at least the prospective ones, and the navigable arc set stays in $\mathcal{S}$ throughout. Induction on $t$ completes the proof. \hfill  $\square$

%% file: sections/architecture.tex
\section{Proposed Architecture}\label{sec:arch}
The road controllers share a single centralized model, a transformer \cite{ext:vaswani2017attention}. The name of its attention mechanism, \emph{Global Endpoint Relation-Aware Spatio-Temporal Self-Attention} (GERASTSA), lists its properties. It is \emph{global}: every token may attend to every road at its own and every earlier time step, not only to its neighbors in the controller graph. It is \emph{endpoint relation-aware}: a learned bias, selected by how two roads are related through their endpoints, is added to every attention score, in the spirit of relation-aware attention \cite{ext:shaw2018relative} and graph transformers \cite{ext:ying2021graphormer}. It is \emph{spatio-temporal}: its \emph{tokens} are road--time pairs, so space and time are attended to jointly, and the GERASTSA transformer adds a spatial embedding of every road and a temporal embedding of every window position to the projected input. The controller of a road is the shared model read out at that road's tokens: the regressor of Section~\ref{sec:exp:pretraining} and the heads of Section~\ref{sec:exp:fine-tuning} act on the per-road outputs.

The graphs $G$ and $L (G) = (\mathcal{R}, \mathcal{E})$ are those of Section~\ref{sec:problem}. The model processes the roads in the order of a fixed bijective \emph{road indexing} $\iota$, with \emph{road index} $i$ for road $\iota (i)$. Its hyperparameters are the input, output, model, and feedforward widths $F_{\mathrm{X}}$, $F_{\mathrm{Y}}$, $F_{\mathrm{Z}}$, and $F_{\mathrm{F}}$; the numbers $H$ of heads, each of \emph{head width} $F_{\mathrm{H}}$, and $N$ of layers; the maximum window length $W_{\max}$; the maximum relation distance $\Lambda_{\max}$; the maximum junction degree $\Gamma_{\max}$; and the stabilizer $\varepsilon_{\mathrm{LN}}$ of layer normalization.
\begin{equation}\label{eq:hyperparameters}
\begin{aligned}
& F_{\mathrm{X}}, F_{\mathrm{Y}}, F_{\mathrm{Z}}, F_{\mathrm{F}}, H, N, W_{\max}, \Gamma_{\max} \in \mathbb{Z}_{\geq 1}; \\
& \Lambda_{\max} \in \mathbb{Z}_{\geq 2}; \quad \varepsilon_{\mathrm{LN}} \in \mathbb{R}_{> 0}; \\
& H \mid F_{\mathrm{Z}}; \quad F_{\mathrm{H}} = \frac{F_{\mathrm{Z}}}{H}; \quad R = |\mathcal{R}|; \\
& \iota : \{1, \dots, R\} \to \mathcal{R}\ \mathrm{bijective};
\end{aligned}
\end{equation}

Bold symbols are learnable, tensor indices start at $1$, and, just as $(t)$ marks time step $t$, the superscripts $[n]$ and $\langle k \rangle$ mark layer $n$ and sample $k$. A product with a matrix, as in $X \boldsymbol{W}_{\mathrm{Q}}$, acts on the last axis; $\top$ swaps the last two axes, and a product of two tensors, as in $Q' K^{\prime \top}$, multiplies their last two axes for every index of the leading axes. A tensor added to one of higher order is repeated along the missing axes, and $\operatorname{Softmax}$ acts on the last axis, with $\exp (- \infty) = 0$. Layer normalization $\operatorname{LN}$ \cite{ext:ba2016layernorm} and the SwiGLU feedforward network $\operatorname{FFN}$ \cite{ext:shazeer2020glu}, with biases in all its linear maps, act on every token separately and receive their hyperparameters after a semicolon.

\subsection{GERASTSA Block}\label{sec:arch:gerastsa}
A GERASTSA block is the attention mechanism on its own. Its input consists of $B$ \emph{samples} of $W$ \emph{window positions}, ordered from past to present, with one token $X_{k, w, i} \in \mathbb{R}^{F_{\mathrm{Z}}}$ per sample $k$, position $w$, and road index $i$. An \emph{attention mask} $M$ allows ($0$) or forbids ($- \infty$) attention between tokens of the same sample, and every token must be allowed to attend to some token.
\begin{equation}\label{eq:gerastsa-input}
\begin{aligned}
& B, W \in \mathbb{Z}_{\geq 1}; \quad X \in \mathbb{R}^{B \times W \times R \times F_{\mathrm{Z}}}; \\
& M \in {\{0, - \infty\}}^{B \times W \times R \times W \times R}; \\
& \exists (w', i') \in \{1, \dots, W\} \times \{1, \dots, R\} : \\
&\ M_{k, w, i, w', i'} = 0\ \forall k \in \{1, \dots, B\} \\
&\ \forall (w, i) \in \{1, \dots, W\} \times \{1, \dots, R\};
\end{aligned}
\end{equation}

Within a sample, attention runs over all $W R$ tokens. The \emph{token index map} $\theta$ flattens window positions and road indices, and the \emph{head-feature index map} $\sigma$ places the features of each head within the model width.
\begin{equation}\label{eq:index-maps}
\begin{aligned}
& \theta : \{1, \dots, W\} \times \{1, \dots, R\} \to \{1, \dots, W R\}; \\
& \theta (w, i) = (w - 1) R + i; \\
& \sigma : \{1, \dots, H\} \times \{1, \dots, F_{\mathrm{H}}\} \to \{1, \dots, F_{\mathrm{Z}}\}; \\
& \sigma (h, f) = (h - 1) F_{\mathrm{H}} + f;
\end{aligned}
\end{equation}

Roads are related through their endpoints. The \emph{head junction} $\eta (r, d)$ is the junction into which road $r$ carries traffic in direction $d$. The \emph{hop set} $\mathcal{N}_{q} (r)$ holds the roads reached from $r$ by $q$ hops in $L (G)$, and the \emph{hop distance} $\lambda (r, r')$ is the fewest hops that reach $r'$; it is finite because $L (G)$ is connected, as $G$ is.
\begin{equation}\label{eq:endpoints}
\begin{aligned}
& \eta : \mathcal{R} \times \mathcal{D} \to \mathcal{J}; \quad \eta (r, d) = \begin{cases}
j_{r}^{+} & \mathrm{if}\ d = \mathrm{fwd} \\
j_{r}^{-} & \mathrm{if}\ d = \mathrm{bwd}
\end{cases}; \\
& \mathcal{N}_{q} : \mathcal{R} \to 2^{\mathcal{R}}, \\
& \mathcal{N}_{q} (r) = \\
&\ \begin{cases}
\{r\} & \mathrm{if}\ q = 0 \\
\bigcup_{r'' \in \mathcal{N}_{q - 1} (r)} \mleft\{r' \in \mathcal{R} \mathrel{}\middle|\mathrel{} \{r'', r'\} \in \mathcal{E}\mright\} & \mathrm{otherwise}
\end{cases} \\
&\ \forall q \in \mathbb{Z}_{\geq 0}; \\
& \lambda : \mathcal{R}^{2} \to \mathbb{Z}_{\geq 0}; \\
& \lambda (r, r') = \min \mleft\{q \in \mathbb{Z}_{\geq 0} \mathrel{}\middle|\mathrel{} r' \in \mathcal{N}_{q} (r)\mright\};
\end{aligned}
\end{equation}

\begin{figure}[tb]
\centering
\includegraphics{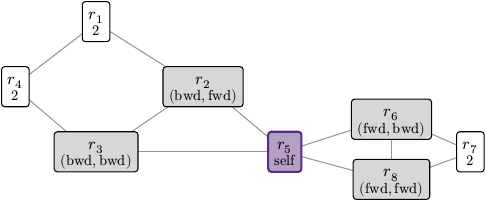}
\caption{Relations $\kappa (r_{5}, \cdot)$ on the controller graph of the example network, with $\Lambda_{\max} = 2$ (heavy border: $r_{5}$; shaded: roads adjacent to $r_{5}$; white: roads at hop distance at least $2$). Roads adjacent to $r_{5}$, with $\omega (r_{5}, \mathrm{fwd}) = (j_{3}, j_{5})$, are told apart by the directions in which the two roads flow into their shared junction, e.g. $r_{2}$, with $\omega (r_{2}, \mathrm{fwd}) = (j_{2}, j_{3})$, flows into $j_{3}$ forward while $r_{5}$ flows into it backward. Each relation $\kappa (r_{5}, r)$ selects the learned relational bias $\boldsymbol{B}_{\mathrm{R}, h, \xi (\kappa (r_{5}, r))}$ of every head $h$.}
\label{fig:kappa}
\end{figure}

The \emph{relation function} $\kappa$ maps every ordered pair of roads to a relation in the \emph{relation set} $\mathcal{K}$. A road has the \emph{self relation} to itself. Two adjacent roads share exactly one junction, since $G$ is simple, and their \emph{adjacency relation} is the unique pair of directions in which both flow into it; for instance, $(\mathrm{fwd}, \mathrm{bwd})$ means that $r$ flows into the shared junction forward and $r'$ backward. Any other pair has the hop distance, capped at $\Lambda_{\max}$, as its \emph{distance relation}, much like the shortest-path spatial encoding of graph transformers \cite{ext:ying2021graphormer} and the clipped relative distances of \cite{ext:shaw2018relative}, so $|\mathcal{K}| = \Lambda_{\max} + 4$. The \emph{relation index} $\xi$ enumerates the relations, and the \emph{relation index matrix} $\Xi$ holds the relation indices of all pairs of road indices (Fig.~\ref{fig:kappa}).
\begin{equation}\label{eq:relations}
\begin{aligned}
& \mathcal{K} = \{\mathrm{self}\} \cup \mathcal{D}^{2} \cup \{2, \dots, \Lambda_{\max}\}; \quad \kappa : \mathcal{R}^{2} \to \mathcal{K}; \\
& \kappa (r, r') = \begin{cases}
\mathrm{self} & \mathrm{if}\ r = r' \\
(d, d') & \mathrm{if}\ \{r, r'\} \in \mathcal{E} \\
\min (\lambda (r, r'), \Lambda_{\max}) & \mathrm{otherwise}
\end{cases} \\
&\ \mathrm{where}\ \eta (r, d) = \eta (r', d'); \\
& \xi : \mathcal{K} \to \{1, \dots, |\mathcal{K}|\}\ \mathrm{bijective}; \\
& \Xi \in {\{1, \dots, |\mathcal{K}|\}}^{R \times R}; \\
& \Xi_{i, i'} = \xi (\kappa (\iota (i), \iota (i')))\ \forall (i, i') \in {\{1, \dots, R\}}^{2};
\end{aligned}
\end{equation}

Every token is projected to a \emph{query}, a \emph{key}, and a \emph{value}.
\begin{equation}\label{eq:projections}
\begin{aligned}
& \boldsymbol{W}_{\mathrm{Q}}, \boldsymbol{W}_{\mathrm{K}}, \boldsymbol{W}_{\mathrm{V}} \in \mathbb{R}^{F_{\mathrm{Z}} \times F_{\mathrm{Z}}}; \quad \boldsymbol{b}_{\mathrm{Q}}, \boldsymbol{b}_{\mathrm{K}}, \boldsymbol{b}_{\mathrm{V}} \in \mathbb{R}^{F_{\mathrm{Z}}}; \\
& Q = X \boldsymbol{W}_{\mathrm{Q}} + \boldsymbol{b}_{\mathrm{Q}},\ K = X \boldsymbol{W}_{\mathrm{K}} + \boldsymbol{b}_{\mathrm{K}}, \\
& V = X \boldsymbol{W}_{\mathrm{V}} + \boldsymbol{b}_{\mathrm{V}}\ \in \mathbb{R}^{B \times W \times R \times F_{\mathrm{Z}}};
\end{aligned}
\end{equation}
The projections are split into the $H$ heads and flattened into token sequences.
\begin{equation}\label{eq:heads}
\begin{aligned}
& Q', K', V' \in \mathbb{R}^{B \times H \times W R \times F_{\mathrm{H}}}; \\
& Q'_{k, h, \theta (w, i), f} = Q_{k, w, i, \sigma (h, f)}, \\
& K'_{k, h, \theta (w, i), f} = K_{k, w, i, \sigma (h, f)}, \\
& V'_{k, h, \theta (w, i), f} = V_{k, w, i, \sigma (h, f)} \\
&\ \forall k \in \{1, \dots, B\} \\
&\ \forall (h, f) \in \{1, \dots, H\} \times \{1, \dots, F_{\mathrm{H}}\} \\
&\ \forall (w, i) \in \{1, \dots, W\} \times \{1, \dots, R\};
\end{aligned}
\end{equation}

Each head learns one \emph{relational bias} per relation, collected in $\boldsymbol{B}_{\mathrm{R}}$. The bias depends only on the roads of two tokens, so a relation is treated alike at every time lag; temporal order enters only through the mask and the temporal embedding of Section~\ref{sec:arch:transformer}. Bias and mask are flattened like the tokens.
\begin{equation}\label{eq:bias-mask}
\begin{aligned}
& \boldsymbol{B}_{\mathrm{R}} \in \mathbb{R}^{H \times |\mathcal{K}|}; \\
& B'_{\mathrm{R}} \in \mathbb{R}^{H \times W R \times W R}; \quad M' \in {\{0, - \infty\}}^{B \times W R \times W R}; \\
& B'_{\mathrm{R}, h, \theta (w, i), \theta (w', i')} = \boldsymbol{B}_{\mathrm{R}, h, \Xi_{i, i'}}\ \forall h \in \{1, \dots, H\} \\
&\ \forall (w, w') \in {\{1, \dots, W\}}^{2}\ \forall (i, i') \in {\{1, \dots, R\}}^{2}; \\
& M'_{k, \theta (w, i), \theta (w', i')} = M_{k, w, i, w', i'}\ \forall k \in \{1, \dots, B\} \\
&\ \forall (w, w') \in {\{1, \dots, W\}}^{2}\ \forall (i, i') \in {\{1, \dots, R\}}^{2};
\end{aligned}
\end{equation}

\begin{figure}[tb]
\centering
\includegraphics{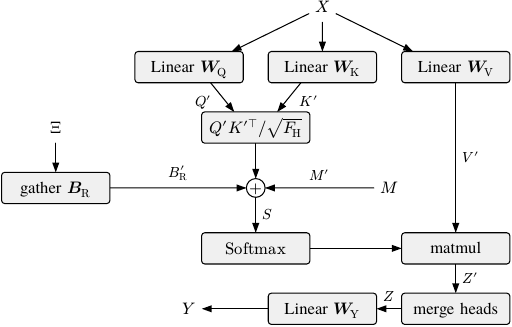}
\caption{The GERASTSA block. Queries, keys, and values are split into $H$ heads and flattened over the $W R$ road--time tokens of each sample. The relational bias $B'_{\mathrm{R}}$ is gathered from the learned table $\boldsymbol{B}_{\mathrm{R}}$ through the relation index matrix $\Xi$ and shared across all window positions and samples, and the flattened mask $M'$ forbids attention as specified by $M$, which in the transformer is the mask of \eqref{eq:mask}.}
\label{fig:gerastsa}
\end{figure}

The attention scores add the relational bias, shared by all samples, and the mask, shared by all heads, to the scaled dot products of queries and keys \cite{ext:vaswani2017attention}. By \eqref{eq:gerastsa-input}, the attention weights are well defined, and they vanish exactly where the mask forbids attention.
\begin{equation}\label{eq:attention}
\begin{aligned}
& S = \frac{Q' K^{\prime \top}}{\sqrt{F_{\mathrm{H}}}} + B'_{\mathrm{R}} + M' \\
&\ \in {(\mathbb{R} \cup \{- \infty\})}^{B \times H \times W R \times W R}; \\
& Z' = \operatorname{Softmax} (S) V'\ \in \mathbb{R}^{B \times H \times W R \times F_{\mathrm{H}}};
\end{aligned}
\end{equation}
Merging the heads
\begin{equation}\label{eq:merge-heads}
\begin{aligned}
& Z \in \mathbb{R}^{B \times W \times R \times F_{\mathrm{Z}}}; \\
& Z_{k, w, i, \sigma (h, f)} = Z'_{k, h, \theta (w, i), f} \\
&\ \forall k \in \{1, \dots, B\} \\
&\ \forall (h, f) \in \{1, \dots, H\} \times \{1, \dots, F_{\mathrm{H}}\} \\
&\ \forall (w, i) \in \{1, \dots, W\} \times \{1, \dots, R\};
\end{aligned}
\end{equation}
and a learnable output projection
\begin{equation}\label{eq:gerastsab}
\begin{aligned}
& \boldsymbol{W}_{\mathrm{Y}} \in \mathbb{R}^{F_{\mathrm{Z}} \times F_{\mathrm{Z}}}; \quad \boldsymbol{b}_{\mathrm{Y}} \in \mathbb{R}^{F_{\mathrm{Z}}}; \\
& Y = Z \boldsymbol{W}_{\mathrm{Y}} + \boldsymbol{b}_{\mathrm{Y}}\ \in \mathbb{R}^{B \times W \times R \times F_{\mathrm{Z}}}; \\
& \operatorname{GERASTSAB} (X, M) = Y;
\end{aligned}
\end{equation}
yield the output of the block (Fig.~\ref{fig:gerastsa}). The symbols of the data flow in \eqref{eq:gerastsa-input} to \eqref{eq:gerastsab} are local to the block.

\subsection{GERASTSA Transformer}\label{sec:arch:transformer}
The GERASTSA transformer maps $B$ histories of the network to one output vector per sample, road, and window position. Sample $k$ ends at a time step $t_{k} \in \breve{\mathcal{T}}$ and holds the input features, defined in Section~\ref{sec:exp}, of all roads at the last $W_{k}$ time steps up to $t_{k}$, ordered from past to present.
\begin{equation}\label{eq:transformer-input}
\begin{aligned}
& B \in \mathbb{Z}_{\geq 1}; \\
& t_{k} \in \breve{\mathcal{T}},\ W_{k} = \min (t_{k} + 1, W_{\max}), \\
& X^{\langle k \rangle} \in \mathbb{R}^{W_{k} \times R \times F_{\mathrm{X}}}\ \forall k \in \{1, \dots, B\};
\end{aligned}
\end{equation}
The samples are left-padded with zeros to the longest window length $W$, so that their last time steps align at position $W$. Position $w$ of sample $k$ is \emph{valid} if $w > W - W_{k}$, in which case it holds time step $t_{k} - (W - w)$, and \emph{padding} otherwise.
\begin{equation}\label{eq:padding}
\begin{aligned}
& W = \max_{k \in \{1, \dots, B\}} W_{k}; \quad X \in \mathbb{R}^{B \times W \times R \times F_{\mathrm{X}}}; \\
& X_{k, w, i, f} = \begin{cases}
X_{w - W + W_{k}, i, f}^{\langle k \rangle} & \mathrm{if}\ w > W - W_{k} \\
0 & \mathrm{otherwise}
\end{cases} \\
&\ \forall k \in \{1, \dots, B\}\ \forall w \in \{1, \dots, W\} \\
&\ \forall i \in \{1, \dots, R\}\ \forall f \in \{1, \dots, F_{\mathrm{X}}\};
\end{aligned}
\end{equation}
The symbols of the data flow in \eqref{eq:transformer-input} to \eqref{eq:gerastsat} are local to the transformer.

The attention mask is \emph{causal}: a token may attend to all roads at the valid positions up to its own. It may also attend to its own position, so that padding tokens satisfy \eqref{eq:gerastsa-input}.
\begin{equation}\label{eq:mask}
\begin{aligned}
& M \in {\{0, - \infty\}}^{B \times W \times R \times W \times R}; \\
& M_{k, w, i, w', i'} = \\
&\ \begin{cases}
0 & \mathrm{if}\ W - W_{k} < w' \leq w \lor w' = w \\
- \infty & \mathrm{otherwise}
\end{cases} \\
&\ \forall k \in \{1, \dots, B\}\ \forall (w, w') \in {\{1, \dots, W\}}^{2} \\
&\ \forall (i, i') \in {\{1, \dots, R\}}^{2};
\end{aligned}
\end{equation}

The \emph{junction degree} $\gamma (r, d)$ is the number of roads that meet at the head junction of road $r$ in direction $d$. The \emph{spatial embedding} $E'_{\mathrm{S}}$ of a road sums, over both directions, the rows that its junction degrees, capped at $\Gamma_{\max}$, select from the learnable tables $\boldsymbol{E}_{\mathrm{S}}^{d}$, a directional variant of the degree-based centrality encoding of \cite{ext:ying2021graphormer}.
\begin{equation}\label{eq:spatial-embedding}
\begin{aligned}
& \gamma : \mathcal{R} \times \mathcal{D} \to \mathbb{Z}_{\geq 1}; \\
& \gamma (r, d) = \mleft|\mleft\{r' \in \mathcal{R} \mathrel{}\middle|\mathrel{} \eta (r, d) \in r'\mright\}\mright|; \\
& \boldsymbol{E}_{\mathrm{S}}^{d} \in \mathbb{R}^{\Gamma_{\max} \times F_{\mathrm{Z}}}\ \forall d \in \mathcal{D}; \quad E'_{\mathrm{S}} \in \mathbb{R}^{R \times F_{\mathrm{Z}}}; \\
& E'_{\mathrm{S}, i, f} = \sum_{d \in \mathcal{D}} \boldsymbol{E}_{\mathrm{S}, \min (\gamma (\iota (i), d), \Gamma_{\max}), f}^{d} \\
&\ \forall i \in \{1, \dots, R\}\ \forall f \in \{1, \dots, F_{\mathrm{Z}}\};
\end{aligned}
\end{equation}

The \emph{temporal embedding} $E'_{\mathrm{T}}$ gives position $w$ row $W_{\max} - (W - w)$ of the learnable table $\boldsymbol{E}_{\mathrm{T}}$, a row determined by how many time steps $w$ precedes the last time step. The last time step thus always receives row $W_{\max}$, however much its sample is padded.
\begin{equation}\label{eq:temporal-embedding}
\begin{aligned}
& \boldsymbol{E}_{\mathrm{T}} \in \mathbb{R}^{W_{\max} \times F_{\mathrm{Z}}}; \quad E'_{\mathrm{T}} \in \mathbb{R}^{W \times F_{\mathrm{Z}}}; \\
& E'_{\mathrm{T}, w, f} = \boldsymbol{E}_{\mathrm{T}, W_{\max} - W + w, f} \\
&\ \forall w \in \{1, \dots, W\}\ \forall f \in \{1, \dots, F_{\mathrm{Z}}\};
\end{aligned}
\end{equation}

\begin{figure}[tb]
\centering
\includegraphics{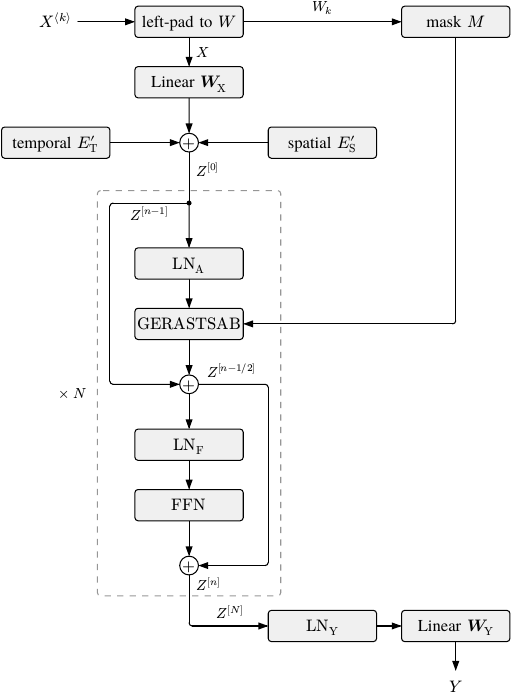}
\caption{The GERASTSA transformer. The histories $X^{\langle k \rangle}$ are left-padded to the longest window length $W$, and the mask $M$ of \eqref{eq:mask} is built from their window lengths $W_{k}$. The embedded input $Z^{[0]}$ passes through $N$ pre-norm layers, each mapping $Z^{[n - 1]}$ to $Z^{[n]}$ with a GERASTSA block and a SwiGLU feedforward network in residual connections; a final layer normalization and a linear map produce the output $Y$.}
\label{fig:transformer}
\end{figure}

A linear map lifts the input to the model width, and the temporal and spatial embeddings are added at every road and every position, respectively.
\begin{equation}\label{eq:input-embedding}
\begin{aligned}
& \boldsymbol{W}_{\mathrm{X}} \in \mathbb{R}^{F_{\mathrm{X}} \times F_{\mathrm{Z}}}; \quad \boldsymbol{b}_{\mathrm{X}} \in \mathbb{R}^{F_{\mathrm{Z}}}; \\
& Z^{[0]} = X \boldsymbol{W}_{\mathrm{X}} + \boldsymbol{b}_{\mathrm{X}} + E'_{\mathrm{T}} + E'_{\mathrm{S}}\ \in \mathbb{R}^{B \times W \times R \times F_{\mathrm{Z}}};
\end{aligned}
\end{equation}
The transformer stacks $N$ pre-norm layers \cite{ext:xiong2020layernorm}, each wrapping a GERASTSA block and a feedforward network in residual connections, and $Z^{[n - 1 / 2]}$ is the residual stream halfway through layer $n$.
\begin{equation}\label{eq:layers}
\begin{aligned}
& Z^{[n - 1 / 2]} = Z^{[n - 1]} \\
&\ {}+ \operatorname{GERASTSAB}^{[n]} \mleft(\operatorname{LN}_{\mathrm{A}}^{[n]} (Z^{[n - 1]}; \varepsilon_{\mathrm{LN}}), M\mright), \\
& Z^{[n]} = Z^{[n - 1 / 2]} \\
&\ {}+ \operatorname{FFN}^{[n]} \mleft(\operatorname{LN}_{\mathrm{F}}^{[n]} (Z^{[n - 1 / 2]}; \varepsilon_{\mathrm{LN}}); F_{\mathrm{F}}\mright) \\
&\ \in \mathbb{R}^{B \times W \times R \times F_{\mathrm{Z}}}\ \forall n \in \{1, \dots, N\};
\end{aligned}
\end{equation}
A final layer normalization and a linear map produce the output (Fig.~\ref{fig:transformer}).
\begin{equation}\label{eq:gerastsat}
\begin{aligned}
& \boldsymbol{W}_{\mathrm{Y}} \in \mathbb{R}^{F_{\mathrm{Z}} \times F_{\mathrm{Y}}}; \quad \boldsymbol{b}_{\mathrm{Y}} \in \mathbb{R}^{F_{\mathrm{Y}}}; \\
& Y = \operatorname{LN}_{\mathrm{Y}} (Z^{[N]}; \varepsilon_{\mathrm{LN}}) \boldsymbol{W}_{\mathrm{Y}} + \boldsymbol{b}_{\mathrm{Y}}\ \in \mathbb{R}^{B \times W \times R \times F_{\mathrm{Y}}}; \\
& \operatorname{GERASTSAT} (X^{\langle 1 \rangle}, \dots, X^{\langle B \rangle}) = Y;
\end{aligned}
\end{equation}

Valid tokens never attend to padding tokens, every operation acts on each sample or token separately, and the temporal embedding is aligned to the last time step. Hence, the outputs of a sample at its valid positions are those it would receive if processed alone, regardless of padding and of the other samples; the outputs at padding positions are ignored.

No parameter belongs to a particular road: the same weights act on the tokens of all roads, which the model tells apart only by their input features, $\Xi$, and $E'_{\mathrm{S}}$. Re-indexing the roads therefore permutes the outputs accordingly, and the output of each road does not depend on $\iota$. Since no parameter shape depends on $R$ or on $G$, the model can also be applied to other road networks, although our experiments use a single one.

%% file: sections/experimental-design.tex
\section{Experimental Design}\label{sec:exp}
\subsection{Environment}\label{sec:exp:environment}
We simulate traffic with a vectorized mesoscopic traffic simulator that runs a batch of \emph{environments} in parallel. Each environment simulates a sequence of \emph{episodes}, single runs of the problem over the time steps $\breve{\mathcal{T}}$, and starts the next one automatically once an episode reaches the horizon $T$. All vehicles are homogeneous CAVs that exchange information with the road infrastructure through vehicle-to-infrastructure communication. The simulator implements the problem of Section~\ref{sec:problem}: it applies the joint action of the road controllers, resolves the lane proposals with Algorithm~\ref{alg:rejection}, and completes the clearing of a road once the vehicles on its clearing lanes have left.

Our network is a $6 \times 6$ grid of junctions with $60$ roads between horizontally and vertically adjacent junctions, every road oriented forward eastward or southward. The roads of each row and column have $2$, $4$, or $6$ lanes, with nominal speed limits of $40$, $60$, and $70$ km/h, respectively, so that $30$ roads have two lanes, $20$ four, and $10$ six, $200$ lanes in total. The blocks lengthen from $300$ to $350$ m near the north-west corner to $650$ m near the south-east corner. Every road of the grid lies on a cycle, so the network has no bridge roads and the bridge indicator of Table~\ref{tab:features} is zero throughout; we keep the feature so that the model applies to other networks. A commercial zone occupies the north-west corner of the grid and a residential zone the south-east corner, and on top of a background demand, the commutes between them cause a morning rush hour toward the commercial zone, peaking at 08:00, and an evening rush hour back, peaking at 17:30. An episode simulates one day, from midnight to midnight, with a decision step every $600$ s, so $T = 144$. In fine-tuning and evaluation, every episode starts with the lanes of each road split evenly between its two directions and every speed limit nominal.

Road $r$ has length $D_{r}$, and each of its lanes holds at most $C_{r}$ vehicles, its \emph{per-lane capacity}; the hyperparameter $L_{\max}$ bounds the total lanes of every road. Table~\ref{tab:hyperparameters} lists this and all other hyperparameters of our experiments. The simulator describes the traffic in each direction of a road by three quantities, computed over the active lanes of the direction. The \emph{congestion} $c_{r, d}^{(t)}$ is the number of vehicles on the most-occupied active lane, and the \emph{density} $\bar{c}_{r, d}^{(t)}$ the mean number of vehicles per active lane, both relative to $C_{r}$. The \emph{normalized velocity} $v_{r, d}^{(t)}$ is the mean velocity of the vehicles divided by the nominal speed limit. All three are $0$ for a direction without active lanes, and the normalized velocity is $1$ for a direction with active lanes but no vehicles, which flows freely.

Vehicles route themselves with knowledge of the current traffic: the cost of a route accounts for the congestion and the mean velocity of every road direction along it. Whenever a vehicle reaches a junction, it checks two conditions and, if either holds, reroutes from that junction along an optimal route; otherwise, it keeps its route. The first is that the next road of its remaining route has no open lane in the direction of the route, as when that direction has been closed. The second is that an optimal route costs less than $1 - \varepsilon_{\mathrm{R}}$ times the remaining route, where $\varepsilon_{\mathrm{R}}$ is the \emph{rerouting threshold}. We chose the threshold arbitrarily; it only keeps vehicles from oscillating between routes of nearly equal cost. The first condition never strands a vehicle: since the navigable graph is strongly connected at every time step (Proposition~\ref{prop:connectivity}), a vehicle that must reroute because a direction has been closed always has a route to the junction of its destination.

\begin{table}[tb]
\centering
\caption{Hyperparameters of the experiments.}\label{tab:hyperparameters}
\footnotesize
\begin{tabular}{@{}llr@{}}
\toprule
\textbf{Symbol} & \textbf{Hyperparameter} & \textbf{Value} \\
\midrule
\multicolumn{3}{l}{\emph{Environment}} \\
$L_{\max}$ & Maximum total lanes & $6$ \\
$\varepsilon_{\mathrm{R}}$ & Rerouting threshold & $0.1$ \\
\midrule
\multicolumn{3}{l}{\emph{GERASTSA transformer}} \\
$F_{\mathrm{Y}}$ & Output width & $128$ \\
$F_{\mathrm{Z}}$ & Model width & $128$ \\
$F_{\mathrm{F}}$ & Feedforward width & $176$ \\
$H$ & Heads & $8$ \\
$N$ & Layers & $4$ \\
$W_{\max}$ & Maximum window length & $16$ \\
$\Lambda_{\max}$ & Maximum relation distance & $6$ \\
$\Gamma_{\max}$ & Maximum junction degree & $6$ \\
$\varepsilon_{\mathrm{LN}}$ & Layer-normalization stabilizer & $10^{- 5}$ \\
\midrule
\multicolumn{3}{l}{\emph{Pretraining}} \\
$\mathcal{H}$ & Horizon set & $\{1, 3, 6\}$ \\
\midrule
\multicolumn{3}{l}{\emph{Reward}} \\
$\beta_{\mathrm{T}}$ & Trip credit & $600$ s \\
$\alpha_{\mathrm{S}}$ & Own weight & $0$ \\
$\alpha_{\mathrm{N}}$ & Neighbor weight & $0.3$ \\
$\alpha_{\mathrm{G}}$ & Network weight & $0.7$ \\
\midrule
\multicolumn{3}{l}{\emph{Policy}} \\
$\sigma_{\min}$ & Minimum speed standard deviation & $0.05$ \\
$\sigma_{\max}$ & Maximum speed standard deviation & $0.25$ \\
\midrule
\multicolumn{3}{l}{\emph{PPO}} \\
$\gamma_{\mathrm{R}}$ & Discount factor & $0.993$ \\
$\lambda_{\mathrm{A}}$ & GAE parameter & $0.95$ \\
$\varepsilon_{\mathrm{P}}$ & Clipping parameter & $0.2$ \\
$c_{\mathrm{V}}$ & Value coefficient & $0.5$ \\
$c_{\mathrm{L}}$ & Lane entropy coefficient & $10^{- 3}$ \\
$c_{\mathrm{U}}$ & Speed entropy coefficient & $10^{- 3}$ \\
\bottomrule
\end{tabular}
\end{table}

Every road is described by the $18$ input features of Table~\ref{tab:features}. The five features in the first four rows are static, and the others change over time. Per-direction features are listed forward first. The target and active lanes enter once each, since their backward counterparts are $L_{r}$ minus the forward ones.

\begin{table}[tb]
\centering
\caption{Input features of road $r$ at time step $t$.}\label{tab:features}
\footnotesize
\begin{tabular}{@{}llr@{}}
\toprule
\textbf{Feature} & \textbf{Value} & \textbf{Count} \\
\midrule
Road length & $D_{r} / \max_{r' \in \mathcal{R}} D_{r'}$ & 1 \\
Total lanes & $L_{r} / L_{\max}$ & 1 \\
Nominal speed limit & $s_{r, d}^{(0)} / \max_{(r', d') \in \mathcal{R} \times \mathcal{D}} s_{r', d'}^{(0)}$ & 2 \\
Bridge indicator & $\mathds{1} (r \in \mathcal{R}_{\leftrightarrow})$ & 1 \\
\midrule
Congestion & $c_{r, d}^{(t)}$ & 2 \\
Density & $\bar{c}_{r, d}^{(t)}$ & 2 \\
Normalized velocity & $v_{r, d}^{(t)}$ & 2 \\
Active normalized speed limit & $u_{r, d}^{(t)}$ & 2 \\
Open lanes & $\bar{\ell}_{r, d}^{(t)} / L_{r}$ & 2 \\
Target lanes & $\hat{\ell}_{r}^{(t)} / L_{r}$ & 1 \\
Active lanes & $\ell_{r}^{(t)} / L_{r}$ & 1 \\
Clearing indicator & $\mathds{1} \mleft(r \in \mathcal{R}_{\approx}^{(t)}\mright)$ & 1 \\
\bottomrule
\end{tabular}
\end{table}

Concatenating the features yields the \emph{feature vector} $x_{r}^{(t)}$, so the input width of Section~\ref{sec:arch} is $F_{\mathrm{X}} = 18$. A sample ends at a decision step or, for the bootstrap value of Section~\ref{sec:exp:fine-tuning:ppo}, at the final time step $T$, and its $W_{k}$ time steps (Section~\ref{sec:arch:transformer}) lie within its episode.

\subsection{Pretraining}\label{sec:exp:pretraining}
Pretraining is a self-supervised, multi-horizon forecasting task. From the history of the network, the model forecasts the congestion, density, and normalized velocity of every road and direction several steps ahead. As in spatio-temporal traffic forecasting \cite{ext:li2018dcrnn,ext:yu2018stgcn}, the task encourages the model to learn how traffic propagates through the network; unlike there, it also encourages the model to learn how traffic responds to lane allocations and speed limits.

The data come from episodes simulated under a random behavior policy. Each episode starts from randomly drawn target lanes, one-way roads included, that satisfy the strong-connectivity requirement in \eqref{eq:strong-connectivity}, with active lanes equal to target lanes and every active normalized speed limit nominal. At randomly chosen decision steps, a road proposes a random admissible lane count, random normalized speed limits for its open directions, or both; otherwise, it proposes its current target lanes and active normalized speed limits, which leaves them unchanged. Action resolution (Section~\ref{sec:problem:action-resolution}) processes every proposal, so the data also cover rejected and relaxed proposals.

The encoder is a GERASTSA transformer with 561,600 parameters that receives the $18$ features; its maximum window length $W_{\max}$ spans $150$ min of history, so a sample with a full window has $W_{\max} R = 960$ road--time tokens. We write $e_{r}^{(t)}$ for its output at road $r$ and the last window position of a sample ending at time step $t$. Only this position serves as the forecast origin, as in fine-tuning: in every layer, its tokens attend to all roads at all valid positions of the window, so the stacked layers can gather the spatial and temporal information of the whole window there.

For every horizon $h$ in the \emph{horizon set} $\mathcal{H}$, a per-road \emph{regressor} $\operatorname{reg}$ forecasts the \emph{forecast quantities} $y_{r}^{(t)}$, the three traffic quantities of Section~\ref{sec:exp:environment} in both directions of road $r$. Following the direct strategy of multi-step forecasting \cite{ext:bentaieb2012review}, it forecasts each horizon separately rather than feeding its own forecasts back, and it does so as a residual on the current values. Besides the encoder output, it receives the one-hot \emph{horizon indicator} $\mathds{1}_{h}$ and the \emph{action context} $p_{r}^{(t, h)}$, which lists the normalized target lanes and the active normalized speed limits of road $r$ at the time steps $t + 1, \dots, t + h$, padded with zeros to the longest horizon. These result from the actions at the decision steps $t, \dots, t + h - 1$ after action resolution, so the forecasts are conditioned on the actions that take effect in between. A paired superscript $(t, h)$ marks a quantity of the forecast from time step $t$ at horizon $h$. Encoder and regressor are trained jointly to minimize the mean squared error over the six forecast quantities and over all samples of a batch, roads, and horizons whose forecast time step $t + h$ lies within the same episode.
\begin{equation}\label{eq:forecast}
\begin{aligned}
& \hat{y}_{r}^{(t, h)} = y_{r}^{(t)} + \operatorname{reg} \mleft(e_{r}^{(t)}, p_{r}^{(t, h)}, \mathds{1}_{h}\mright) \\
&\ \forall r \in \mathcal{R}\ \forall h \in \mathcal{H};\ \forall t \in \{0, \dots, T - h\}; \\
& J_{\mathrm{F}} = \hat{\mathbb{E}}_{t, r, h} \mleft[\frac{1}{6} \mleft\|\hat{y}_{r}^{(t, h)} - y_{r}^{(t + h)}\mright\|^{2}\mright];
\end{aligned}
\end{equation}
After pretraining, the regressor is discarded, and the encoder initializes the fine-tuning.

\subsection{Fine-Tuning}\label{sec:exp:fine-tuning}
Fine-tuning is an RL task in which the model learns to propose lane counts and normalized speed limits that maximize the return.

\subsubsection{Reward}\label{sec:exp:fine-tuning:reward}
Over decision step $t$, road $r$ completes $m_{r}^{(t)}$ trips and accrues the delay $z_{r}^{(t)}$: the time spent by vehicles waiting for admission at the road and the time lost by vehicles traveling on it below its nominal speed limit. Its reward $g_{r}^{(t)}$ credits each completed trip with $\beta_{\mathrm{T}} > 0$ and charges the delay. The reward that the road optimizes, its \emph{mixed reward} $g_{\mathrm{M}, r}^{(t)}$, combines its own reward, the mean reward of its neighbors in the controller graph, and the mean reward of the whole network with nonnegative weights $\alpha_{\mathrm{S}}$, $\alpha_{\mathrm{N}}$, and $\alpha_{\mathrm{G}}$ that sum to $1$. It resembles the spatially discounted rewards of multi-agent traffic signal control \cite{ext:chu2020marl}, which weight the reward of every agent by a power of its distance, but it keeps a fixed share for the whole network.
\begin{equation}\label{eq:mixed-reward}
\begin{aligned}
& g_{r}^{(t)} = \beta_{\mathrm{T}} m_{r}^{(t)} - z_{r}^{(t)}, \\
& g_{\mathrm{M}, r}^{(t)} = \alpha_{\mathrm{S}} g_{r}^{(t)} + \frac{\alpha_{\mathrm{N}}}{|\mathcal{N}_{1} (r)|} \sum_{r' \in \mathcal{N}_{1} (r)} g_{r'}^{(t)} \\
&\ {}+ \frac{\alpha_{\mathrm{G}}}{R} \sum_{r' \in \mathcal{R}} g_{r'}^{(t)}\ \forall r \in \mathcal{R}\ \forall t \in \mathcal{T};
\end{aligned}
\end{equation}

Because roads may become one-way, we keep the own weight $\alpha_{\mathrm{S}}$ close to zero. Closing a direction sheds its traffic onto other roads: the road's own reward can rise, since fewer vehicles travel on it and accrue delay there, while the network as a whole can lose. With a large own weight, a road could profit from this, which could push the policy toward one-way roads. Weighting the neighbors and the network makes a road bear part of the cost of the traffic it sheds. In our experiments, $\alpha_{\mathrm{S}} = 0$ (Table~\ref{tab:hyperparameters}), so a road's own reward still enters its mixed reward, but only through the network mean, with weight $\alpha_{\mathrm{G}} / R = 0.7 / 60$, since $\mathcal{N}_{1} (r)$ does not contain $r$.

\subsubsection{Policy and Value Networks}\label{sec:exp:fine-tuning:networks}
The actor and the critic are separate networks, each consisting of a GERASTSA transformer and per-road heads, and both transformers are initialized from the pretrained encoder. The heads act on the outputs at the last time step of a sample (Section~\ref{sec:exp:environment}). The critic has a value head that yields the value $V_{r}^{(t)}$ of road $r$. The actor has a lane head and, per direction, two speed heads.

The \emph{lane head} outputs one logit for every lane count from $0$ to $L_{\max}$, the proposed target lanes of the forward direction; the backward direction receives the remaining lanes. Since $L_{\max}$ is a fixed hyperparameter, the lane head has a fixed output size, independent of the road and the network. Lane counts outside the admissible set of \eqref{eq:admissible-lanes} are masked out of the softmax, as in invalid action masking \cite{ext:huang2022masking}, which yields the lane distribution $\pi_{\mathrm{L}, r}^{(t)}$ over the admissible lane counts.

The \emph{speed heads} of direction $d$ output a mean offset and a scale logit. The proposed normalized speed limit follows the speed distribution $\pi_{\mathrm{U}, r, d}^{(t)}$, the truncation to $[0, 1]$ of a normal distribution whose mean is $1$ plus the offset and whose standard deviation the sigmoid of the scale logit places between the bounds $\sigma_{\min}$ and $\sigma_{\max}$. Since the offsets of the untrained speed heads are close to zero, shifting the mean by $1$ keeps speed limits close to nominal early in fine-tuning, which is meant to stabilize its first updates.

During fine-tuning, the proposed lanes and normalized speed limits are sampled from these distributions; during execution, the modes are taken. By \eqref{eq:admissible-speeds}, a direction without open lanes keeps its active normalized speed limit, so its speed heads are ignored. The log-probability of the action $a_{r}^{(t)}$ of road $r$ thus sums the log-probabilities of its proposed lanes and of the proposed normalized speed limits of its \emph{open directions} $\mathcal{O}_{r}^{(t)}$ only.
\begin{equation}\label{eq:log-probability}
\begin{aligned}
& \log \pi_{r}^{(t)} \mleft(a_{r}^{(t)}\mright) = \log \pi_{\mathrm{L}, r}^{(t)} \mleft(\tilde{\ell}_{r}^{(t)}\mright) \\
&\ {}+ \sum_{d \in \mathcal{O}_{r}^{(t)}} \log \pi_{\mathrm{U}, r, d}^{(t)} \mleft(\tilde{u}_{r, d}^{(t)}\mright)\ \forall r \in \mathcal{R}\ \forall t \in \mathcal{T};
\end{aligned}
\end{equation}

\subsubsection{Per-Road PPO}\label{sec:exp:fine-tuning:ppo}
We train actor and critic with PPO \cite{ext:schulman2017ppo}, applied per road. As in independent and multi-agent PPO \cite{ext:dewitt2020ippo,ext:yu2022mappo}, every road acts as an agent with its own probability ratio and its own advantages, while all roads share the parameters. Unlike in these methods, our actor observes the whole network (Section~\ref{sec:prior-work:rl}).

Each update collects a \emph{rollout} of a fixed number of consecutive decision steps from every environment. A rollout may span the end of one episode and the start of the next. The advantages $\hat{A}_{r}^{(t)}$ follow truncated generalized advantage estimation (GAE) \cite{ext:schulman2016gae,ext:schulman2017ppo}, computed from the mixed rewards of the road and its values under the critic $V_{\mathrm{old}, r}^{(t)}$ that collected the data, with discount factor $\gamma_{\mathrm{R}}$ and GAE parameter $\lambda_{\mathrm{A}}$. A rollout \emph{ends} at its last decision step and at the last decision step of every episode it contains; there, the recursion stops and bootstraps from the critic's value at the next time step. Since traffic continues beyond the horizon, the end of an episode is a time-limit truncation rather than a termination \cite{ext:pardo2018timelimits}. The discount factor gives an effective horizon of about $1 / (1 - \gamma_{\mathrm{R}}) \approx 143$ decision steps, roughly one day. The \emph{return estimates} $\hat{V}_{r}^{(t)}$ add the values back to the advantages.
\begin{equation}\label{eq:gae}
\begin{aligned}
& \delta_{r}^{(t)} = g_{\mathrm{M}, r}^{(t)} + \gamma_{\mathrm{R}} V_{\mathrm{old}, r}^{(t + 1)} - V_{\mathrm{old}, r}^{(t)}, \\
& \hat{A}_{r}^{(t)} = \begin{cases}
\delta_{r}^{(t)} & \mathrm{if}\ t\ \mathrm{ends\ the\ rollout} \\
\delta_{r}^{(t)} + \gamma_{\mathrm{R}} \lambda_{\mathrm{A}} \hat{A}_{r}^{(t + 1)} & \mathrm{otherwise}
\end{cases}, \\
& \hat{V}_{r}^{(t)} = \hat{A}_{r}^{(t)} + V_{\mathrm{old}, r}^{(t)}\ \forall r \in \mathcal{R}\ \forall t \in \mathcal{T};
\end{aligned}
\end{equation}

The probability ratio $\zeta_{r}^{(t)}$ compares the probability of the action of road $r$ under the current policy with that under the policy that collected the data.
\begin{equation}\label{eq:ratio}
\begin{aligned}
& \zeta_{r}^{(t)} = \frac{\pi_{r}^{(t)} \mleft(a_{r}^{(t)}\mright)}{\pi_{\mathrm{old}, r}^{(t)} \mleft(a_{r}^{(t)}\mright)}\ \forall r \in \mathcal{R}\ \forall t \in \mathcal{T};
\end{aligned}
\end{equation}
It enters the clipped surrogate objective $J_{\mathrm{P}}$ through the helper function $\chi$, which clips it with the clipping parameter $\varepsilon_{\mathrm{P}}$, and the critic regresses the return estimates through the squared error $J_{\mathrm{V}}$; both average over the decision steps of a minibatch and all roads.
\begin{equation}\label{eq:surrogate}
\begin{aligned}
& \chi : \mathbb{R}_{> 0} \times \mathbb{R} \to \mathbb{R}; \\
& \chi (x, y) = \min (x y, \operatorname{clip} (x, 1 - \varepsilon_{\mathrm{P}}, 1 + \varepsilon_{\mathrm{P}}) y); \\
& J_{\mathrm{P}} = \hat{\mathbb{E}}_{t, r} \mleft[\chi \mleft(\zeta_{r}^{(t)}, \hat{A}_{r}^{(t)}\mright)\mright]; \\
& J_{\mathrm{V}} = \hat{\mathbb{E}}_{t, r} \mleft[{\mleft(V_{r}^{(t)} - \hat{V}_{r}^{(t)}\mright)}^{2}\mright];
\end{aligned}
\end{equation}
As in PPO, the combined objective $J$ adds the mean entropies $\mathbb{H}_{\mathrm{L}}$ and $\mathbb{H}_{\mathrm{U}}$ of the lane and speed distributions as a bonus \cite{ext:mnih2016a3c}, the latter over open directions only, and weights the terms with the coefficients $c_{\mathrm{V}}$, $c_{\mathrm{L}}$, and $c_{\mathrm{U}}$. The objective is maximized.
\begin{equation}\label{eq:fine-tuning-objective}
\begin{aligned}
& J = J_{\mathrm{P}} - c_{\mathrm{V}} J_{\mathrm{V}} + c_{\mathrm{L}} \mathbb{H}_{\mathrm{L}} + c_{\mathrm{U}} \mathbb{H}_{\mathrm{U}};
\end{aligned}
\end{equation}
As the networks are separate, $J_{\mathrm{V}}$ trains only the critic, and the other terms train only the actor.

%% file: sections/results.tex
\section{Results}\label{sec:results}
We evaluate every policy on $32$ days, $8$ episodes for each of $4$ seeds not used in training. A seed and an episode index determine the demand, which is therefore the same under every policy, so that comparisons between policies are paired by day. The learned policies take the modes of their distributions (Section~\ref{sec:exp:fine-tuning:networks}). We report the return, the reward $g_{r}^{(t)}$ summed over all roads and decision steps of a day, in vehicle-hours (veh\textperiodcentered{}h), so that each completed trip counts as $\beta_{\mathrm{T}} = 600$ s of avoided delay; its two components, the delay and the number of completed trips; the mean travel time of a trip from its request to its arrival and the mean admission wait it spends at its origin road; and the number of closing proposals that action resolution rejects. As a safeguard that is not part of the problem of Section~\ref{sec:problem}, the simulator removes every vehicle that has queued for $600$ s, since a gridlocked lane could otherwise block clearing for the rest of the day; we report these removals as well. Values are means and standard errors over the $32$ days, and changes against a reference are means and standard errors of the paired differences.

All variants share the same data and schedule. Pretraining runs for 30,000 steps of $32$ samples on 3,584 days simulated under the behavior policy of Section~\ref{sec:exp:pretraining}, and its validation error is measured on $512$ further days. Fine-tuning runs for $100$ PPO updates, each of which collects $8$ decision steps from each of 1,024 environments, about 5,700 days in total. Before the first update, the environments run under the baseline of Section~\ref{sec:results:learned} for up to one day, so that their clocks are spread uniformly over the day. Every $10$ updates, the greedy policy, which takes the modes, is evaluated on $8$ further days with fixed demand, which yields the learning curves below; their reference is the baseline on the same days. Each variant is trained once with the same seed, so the standard errors reflect the variation across evaluation days but not across training runs.

\subsection{Effect of Pretraining}\label{sec:results:pretraining}
\begin{figure}[tb]
\centering
\includegraphics[width=\linewidth]{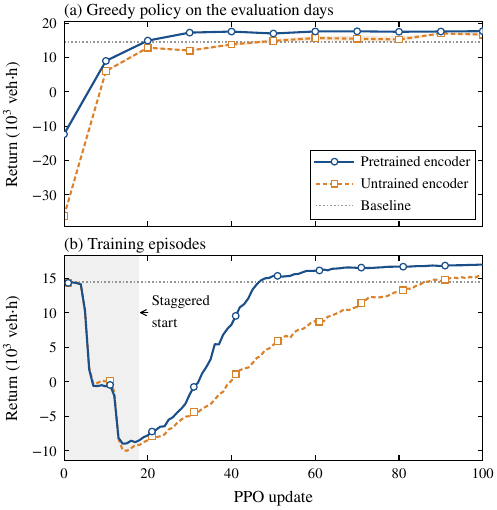}
\caption{Return against the PPO update when fine-tuning from the pretrained encoder and from an untrained one. (a) Greedy policy on the $8$ days evaluated during fine-tuning (shaded: standard error). (b) Mean return of the training episodes that end in each update, under sampled actions; until update $18$ (shaded), these episodes began under the baseline of the warm-up, and the drops at updates $5$ and $6$ and at updates $12$ to $14$ mark the policy taking over the evening and the morning rush hour. Dotted: mean return of the baseline on the days of (a).}
\label{fig:pretraining}
\end{figure}

Fig.~\ref{fig:pretraining} compares fine-tuning from the pretrained encoder with fine-tuning from an untrained, randomly initialized one, under otherwise identical settings. The pretrained encoder forecasts the traffic quantities with a validation MSE of $0.0143$, which is $59 \%$, $70 \%$, and $77 \%$ below that of persistence, the forecast that nothing changes, at the horizons of $10$, $30$, and $60$ min.

Panel (b) of Fig.~\ref{fig:pretraining} must be read with the warm-up in mind. When fine-tuning starts, each environment is in the middle of a day that began under the baseline, and a day spans $18$ updates, so a training episode that ends at update $k \leq 18$ was controlled by the baseline until about $8 k$ decision steps before midnight and, after that, by the early policy, whose untrained heads still propose almost at random. The return of the training episodes thus stays at the level of the baseline while the policy controls only the quiet night, up to update $4$, and falls in two steps as the policy takes over the evening rush hour, at updates $5$ and $6$, and the morning rush hour, at updates $12$ to $14$: its random reallocations cost little at night but cause queues and gridlock under heavy demand. Between the two steps, the return lies near zero, and from update $14$ to $18$, it bottoms out at about \textminus{}9,000 veh\textperiodcentered{}h with the pretrained encoder and \textminus{}10,000 veh\textperiodcentered{}h with the untrained one. This transient stems from the staggered start, not from a collapse of the policy, and the greedy policy shows no such dip (Fig.~\ref{fig:pretraining} (a)). Until about update $15$, the two curves nearly coincide, since most of each finished day was controlled by the baseline after an identical warm-up, and the actor's encoder stays frozen during the first two updates, the first of which trains only the critic. From update $18$ on, panel (b) tracks learning, but under sampled actions and with a lag of about $9$ updates, since an episode that ends at update $k$ was controlled by the policies of updates $k - 17$ to $k$. The run with the pretrained encoder recovers past the reference line of the baseline at update $47$, and the run with the untrained encoder at update $86$; since the training episodes have other demand than the days of that line, these crossings are approximate.

The greedy policy shows the same gap earlier (Fig.~\ref{fig:pretraining} (a)). With the pretrained encoder, it surpasses the baseline, which attains about 14,500 veh\textperiodcentered{}h on the same days, from update $20$ on and varies only between 16,976 and 17,656 veh\textperiodcentered{}h from update $30$; with the untrained encoder, it first surpasses the baseline at update $50$ and varies between 12,007 and 16,983 veh\textperiodcentered{}h from update $30$. On the evaluation days, the final policy from the pretrained encoder attains a return of 17,653 veh\textperiodcentered{}h against 17,112 veh\textperiodcentered{}h from the untrained encoder, which falls short by $3.1 \pm 0.2 \%$, and it incurs $374$ veh\textperiodcentered{}h of delay against $911$ veh\textperiodcentered{}h (Table~\ref{tab:results}). Pretraining thus roughly halves the number of updates needed to surpass the baseline and, after $100$ updates, leaves less than half the delay.

\subsection{Learned Policy}\label{sec:results:learned}
The baseline holds the initial configuration of every episode throughout the day: every road keeps its lanes split evenly between its two directions, and every speed limit stays nominal. Table~\ref{tab:results} lists the metrics of all policies. Against the baseline, the learned policy raises the return by $22.9 \pm 0.3 \%$ and reduces the delay by $89.5 \pm 1.3 \%$, from 3,565 to $374$ veh\textperiodcentered{}h per day. The mean travel time falls from $5.20$ to $3.52$ min, by $32 \%$, and the mean admission wait from $25$ to $3$ s. The learned policy completes $583$ more trips per day, $0.5 \%$ more, exactly as many as the safeguard removes from gridlocked queues under the baseline, and never under the learned policy, so the gain in return stems almost entirely from the delay.

As a second baseline, the \emph{two-way load balancer} reallocates lanes in response to the observed traffic but never makes a road one-way. At each decision step, it takes as the load of each direction of a road its density times its active lanes, i.e., its vehicles in units of $C_{r}$, averaged over the last few decision steps. Unless the road is clearing or carries fewer vehicles than a threshold, it proposes to split the lanes in proportion to the loads of the two directions, rounded and kept between $1$ and $L_{r} - 1$, whenever this split differs from the current target lanes by at least half a lane plus a margin; every speed limit stays nominal. A grid search on $32$ days drawn from the seed of fine-tuning, not from the evaluation seeds, chose the averaging window among $1$, $3$, and $6$ decision steps, the margin among $0$, $0.25$, and $0.5$ lanes, and the threshold among $0.01$, $0.05$, and $0.1$ times $C_{r}$, for the highest mean return. The load balancer raises the return by $7.3 \pm 0.2 \%$ and reduces the delay by $29.2 \pm 0.9 \%$ against the fixed configuration, to 2,524 veh\textperiodcentered{}h, but the safeguard still removes $535$ vehicles per day, and the mean admission wait rises from $25$ to $39$ s. The learned policy attains a $14.5 \%$ higher return than the load balancer, with $85 \%$ less delay.

\begin{table*}[t]
\centering
\caption{Performance on the $32$ evaluation days: means and standard errors over the days, with changes in return paired by day. Load balancer: the heuristic baseline of Section~\ref{sec:results:learned}; full model: the learned policy; untrained encoder: the full model fine-tuned without pretraining; the six rows from lane-only on: the full model restricted at test time (Section~\ref{sec:results:restricted}); the last five rows: the ablations (Section~\ref{sec:results:ablation}). Removed: vehicles removed from gridlocked queues by the safeguard; rejected: closing proposals rejected by action resolution, both per day.}\label{tab:results}
\footnotesize
\setlength{\tabcolsep}{3pt}
\begin{tabular}{@{}lrrrrrrrrr@{}}
\toprule
\textbf{Policy} & \textbf{Return} & \multicolumn{2}{c}{\textbf{Return change (\%)}} & \textbf{Delay} & \textbf{Trips} & \textbf{Travel time} & \textbf{Adm. wait} & \textbf{Removed} & \textbf{Rejected} \\
 & ($10^{3}$ veh\textperiodcentered{}h) & vs. baseline & vs. full model & (veh\textperiodcentered{}h) & ($10^{3}$) & (min) & (s) &  &  \\
\midrule
Baseline & $14.36 \pm 0.04$ &  & $- 18.6 \pm 0.3$ & $3565 \pm 43$ & $107.58 \pm 0.07$ & $5.20 \pm 0.02$ & $25.4 \pm 0.6$ & $583 \pm 23$ & $0$ \\
Load balancer & $15.41 \pm 0.03$ & $+ 7.3 \pm 0.2$ & $- 12.7 \pm 0.2$ & $2524 \pm 29$ & $107.62 \pm 0.07$ & $4.61 \pm 0.02$ & $39.0 \pm 0.7$ & $535 \pm 17$ & $0$ \\
\midrule
Full model & $17.65 \pm 0.01$ & $+ 22.9 \pm 0.3$ &  & $374 \pm 9$ & $108.16 \pm 0.07$ & $3.52 \pm 0.01$ & $3.1 \pm 0.1$ & $0$ & $161 \pm 2$ \\
Untrained encoder & $17.11 \pm 0.03$ & $+ 19.1 \pm 0.3$ & $- 3.1 \pm 0.2$ & $911 \pm 29$ & $108.14 \pm 0.07$ & $3.88 \pm 0.02$ & $6.7 \pm 0.8$ & $13 \pm 4$ & $211 \pm 16$ \\
\midrule
Lane-only & $17.85 \pm 0.01$ & $+ 24.3 \pm 0.3$ & $+ 1.1 \pm 0.1$ & $174 \pm 7$ & $108.16 \pm 0.07$ & $3.410 \pm 0.005$ & $2.8 \pm 0.1$ & $0$ & $154 \pm 1$ \\
Speed-only & $14.07 \pm 0.05$ & $- 2.0 \pm 0.4$ & $- 20.3 \pm 0.3$ & $3846 \pm 50$ & $107.50 \pm 0.07$ & $5.35 \pm 0.03$ & $28.6 \pm 0.6$ & $659 \pm 21$ & $0$ \\
Two-way relaxed & $15.10 \pm 0.03$ & $+ 5.1 \pm 0.3$ & $- 14.5 \pm 0.2$ & $2843 \pm 34$ & $107.66 \pm 0.07$ & $4.82 \pm 0.02$ & $27.5 \pm 0.8$ & $499 \pm 18$ & $0$ \\
Two-way masked & $15.13 \pm 0.05$ & $+ 5.3 \pm 0.3$ & $- 14.3 \pm 0.2$ & $2815 \pm 41$ & $107.66 \pm 0.08$ & $4.81 \pm 0.02$ & $26.1 \pm 0.6$ & $498 \pm 19$ & $0$ \\
Two-way relaxed, lane-only & $15.53 \pm 0.04$ & $+ 8.1 \pm 0.4$ & $- 12.0 \pm 0.2$ & $2419 \pm 41$ & $107.68 \pm 0.07$ & $4.59 \pm 0.02$ & $25.2 \pm 0.9$ & $479 \pm 19$ & $0$ \\
Two-way masked, lane-only & $15.58 \pm 0.03$ & $+ 8.5 \pm 0.2$ & $- 11.7 \pm 0.2$ & $2368 \pm 32$ & $107.68 \pm 0.07$ & $4.56 \pm 0.02$ & $23.9 \pm 0.8$ & $475 \pm 16$ & $0$ \\
\midrule
No relational bias & $16.26 \pm 0.05$ & $+ 13.2 \pm 0.5$ & $- 7.9 \pm 0.3$ & $1742 \pm 54$ & $108.00 \pm 0.07$ & $4.78 \pm 0.03$ & $5.8 \pm 0.2$ & $159 \pm 21$ & $670 \pm 14$ \\
Collapsed relations & $17.47 \pm 0.02$ & $+ 21.7 \pm 0.3$ & $- 1.0 \pm 0.1$ & $547 \pm 18$ & $108.13 \pm 0.07$ & $3.62 \pm 0.01$ & $2.2 \pm 0.1$ & $16 \pm 8$ & $225 \pm 6$ \\
No spatial embedding & $16.15 \pm 0.07$ & $+ 12.4 \pm 0.6$ & $- 8.5 \pm 0.4$ & $1862 \pm 65$ & $108.05 \pm 0.07$ & $4.68 \pm 0.04$ & $11.5 \pm 1.0$ & $87 \pm 24$ & $560 \pm 19$ \\
Undirected spatial emb. & $17.51 \pm 0.03$ & $+ 21.9 \pm 0.3$ & $- 0.8 \pm 0.2$ & $516 \pm 24$ & $108.16 \pm 0.07$ & $3.78 \pm 0.01$ & $2.8 \pm 0.1$ & $0.2 \pm 0.2$ & $171 \pm 4$ \\
No temporal embedding & $17.47 \pm 0.05$ & $+ 21.6 \pm 0.4$ & $- 1.0 \pm 0.3$ & $552 \pm 47$ & $108.14 \pm 0.07$ & $3.86 \pm 0.03$ & $8.0 \pm 1.4$ & $5 \pm 4$ & $383 \pm 9$ \\
\bottomrule
\end{tabular}
\end{table*}

\begin{figure*}[t]
\centering
\includegraphics[width=\linewidth]{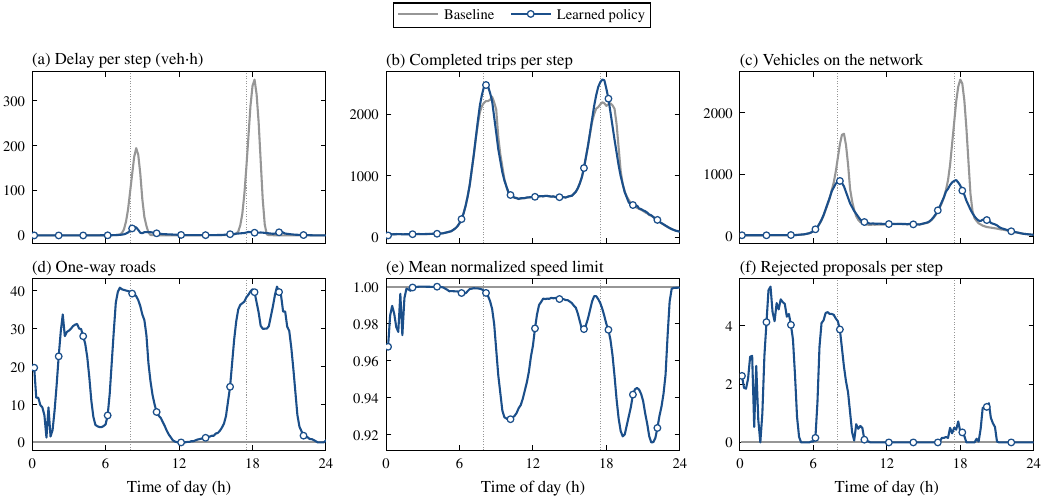}
\caption{Metrics per decision step over the day under the baseline and the learned policy, averaged over the $32$ evaluation days: (a) delay, (b) completed trips, (c) vehicles on the network, (d) one-way roads, (e) mean normalized speed limit over all roads and directions, and (f) closing proposals rejected by action resolution. Dotted verticals mark 08:00 and 17:30, the times shown in Fig.~\ref{fig:congestion}.}
\label{fig:day}
\end{figure*}

Fig.~\ref{fig:day} shows how these differences arise over the day. Under the baseline, $99 \%$ of the delay accrues in the two rush hours, between 06:00 and 11:00 and between 15:00 and 21:00. The delay per decision step peaks at $194$ veh\textperiodcentered{}h around 08:30 and at $347$ veh\textperiodcentered{}h around 18:10, and the vehicles on the network peak at 1,658 and 2,532 at the same times. Under the learned policy, the delay per decision step never exceeds $19$ veh\textperiodcentered{}h, and the vehicles on the network peak at about $900$ in both rush hours, while more trips are completed per decision step near the peaks: the network discharges its demand as it arrives.

\begin{figure*}[t]
\centering
\includegraphics[width=\linewidth]{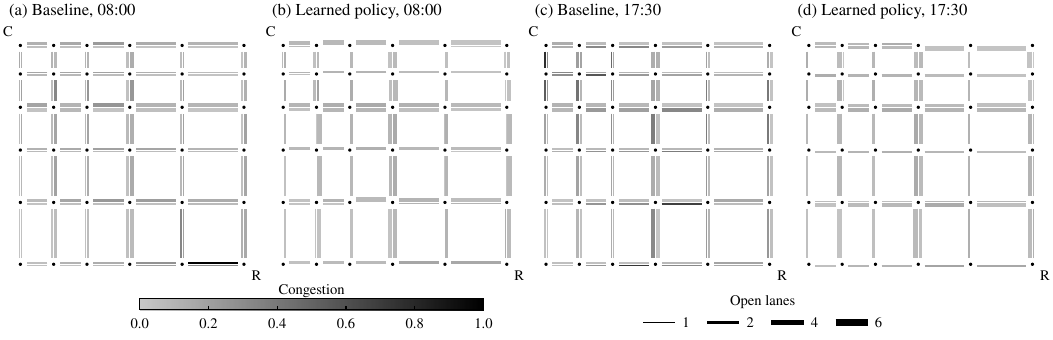}
\caption{Congestion and open lanes of every road and direction at 08:00 and 17:30 under the baseline and the learned policy, averaged over the $32$ evaluation days. Each direction of a road is drawn on its right-hand side, with its luminance giving the congestion and its width the open lanes, so that a one-way road shows a single stroke. North is up; C and R mark the corners of the commercial and the residential zone.}
\label{fig:congestion}
\end{figure*}

Fig.~\ref{fig:congestion} shows where congestion builds up. Under the baseline at 08:00, the directions toward the commercial zone in the north-west are more congested than those away from it, $0.16$ against $0.07$ on average, and the most congested direction, at $0.995$, is the westbound road at the residential corner in the south-east, where morning trips enter the network. At 17:30, congestion is heavier and lies in the directions away from the commercial zone, $0.26$ against $0.08$ on average, with $20$ of them above $0.3$. It is highest on the roads leaving the commercial corner, up to $0.77$, and on roads near the residential corner, up to $0.73$. The learned policy opens on average $155$ of the $200$ lanes of the network toward the commercial zone at 08:00 and $159$ away from it at 17:30, and $30$ and $35$ roads, respectively, are one-way in the peak direction on all evaluation days. No direction then exceeds a congestion of $0.19$.

Over the day, the learned policy makes about $40$ of the $60$ roads one-way during each rush hour and almost none around midday, when only the background demand remains (Fig.~\ref{fig:day} (d)). It hardly lowers speed limits during the rush hours, where the mean normalized speed limit stays above $0.96$, and lowers them slightly after each rush hour, to about $0.92$ to $0.93$ (Fig.~\ref{fig:day} (e)). Action resolution rejects $161$ of its closing proposals per day, at most about $5$ per decision step (Fig.~\ref{fig:day} (f)). Of these, $89$ fall between midnight and 05:00, when the policy also keeps up to $34$ roads one-way with almost no traffic, and $54$ fall between 06:30 and 09:30, at the onset of the morning rush hour; between 16:00 and 19:30, only $4$ are rejected.

\subsection{Restricted Policies}\label{sec:results:restricted}
To see which actions the gain rests on, we evaluate the final full model with part of its control withheld at test time, without retraining. Under \emph{lane-only} control, the policy proposes the lane allocations, and every speed limit stays nominal; under \emph{speed-only} control, it sets the speed limits, and every road keeps its initial lane allocation. Under \emph{two-way} control, the policy may not close a direction: in the \emph{relaxed} variant, each of its proposals of $0$ or $L_{r}$ is replaced by its relaxed proposal as in \eqref{eq:relaxation}, and in the \emph{masked} variant, these two allocations are excluded from the lane distribution before its mode is taken. Two-way control is evaluated both with the policy's speed limits and with nominal ones. These evaluations show what the trained policy retains under each restriction, not what a policy trained under it could reach.

Lane-only control does better than the full model: it raises the return by $1.1 \pm 0.1 \%$ and reduces the delay by $53.5 \pm 2.7 \%$, from $374$ to $174$ veh\textperiodcentered{}h per day. Speed-only control does worse than the baseline, lowering the return by $2.0 \pm 0.4 \%$ and raising the delay by $7.9 \pm 1.5 \%$. Under two-way control, too, the policy's speed limits lower the return: with them, the relaxed and the masked variant raise the return against the baseline by $5.1 \pm 0.3 \%$ and $5.3 \pm 0.3 \%$, and with nominal speed limits by $8.1 \pm 0.4 \%$ and $8.5 \pm 0.2 \%$. All four two-way variants thus fall far short of the $22.9 \%$ of the full model and bracket the $7.3 \%$ of the load balancer; they leave between 2,368 and 2,843 veh\textperiodcentered{}h of delay, and the safeguard removes between $475$ and $499$ vehicles per day.

\subsection{Ablations}\label{sec:results:ablation}
\begin{figure*}[t]
\centering
\includegraphics[width=\linewidth]{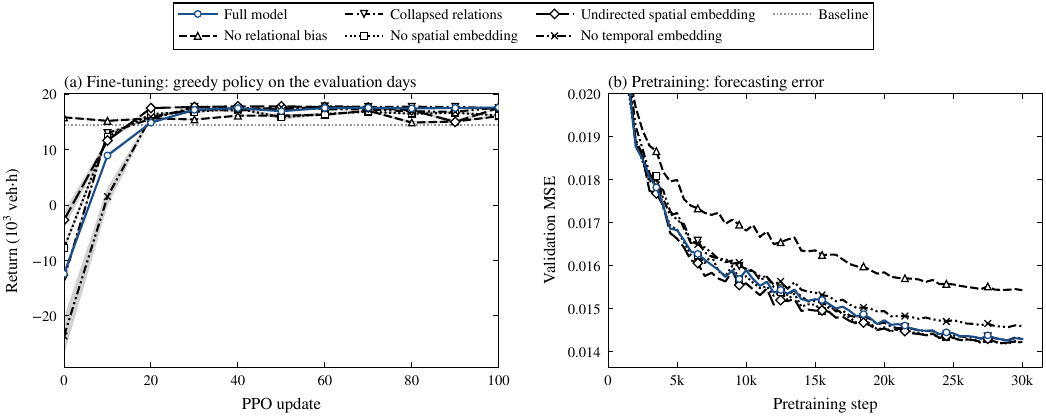}
\caption{Ablations of the GERASTSA transformer against the full model. (a) Return of the greedy policy against the PPO update on the $8$ days evaluated during fine-tuning (shaded: standard error); dotted: mean return of the baseline on the same days. (b) Validation MSE of pretraining against the pretraining step.}
\label{fig:ablations}
\end{figure*}

Each ablation removes one component of the GERASTSA transformer (Section~\ref{sec:arch}) and is pretrained and fine-tuned in the same way as the full model. Fig.~\ref{fig:ablations} compares their learning curves and pretraining errors with those of the full model, and Table~\ref{tab:results} their final performance.

\begin{itemize}
\item \emph{No relational bias:} $B'_{\mathrm{R}}$ is dropped from the attention scores in \eqref{eq:attention}, so attention is not informed by any relation between roads, and roads are told apart only by their input features and the spatial embedding.

\item \emph{Collapsed relations:} the four adjacency relations in $\mathcal{D}^{2}$ are collapsed into a single 1-hop relation, so that $|\mathcal{K}| = \Lambda_{\max} + 1$ and the bias depends on the hop distance alone, no longer on the directions in which adjacent roads flow into their shared junction.

\item \emph{No spatial embedding:} $E'_{\mathrm{S}}$ is dropped from \eqref{eq:input-embedding}, so the junction degrees of a road no longer enter the model directly, and roads are told apart only by their input features and the relational bias.

\item \emph{Undirected spatial embedding:} instead of embedding the junction degree of a road in each direction through the directional tables $\boldsymbol{E}_{\mathrm{S}}^{d}$ of \eqref{eq:spatial-embedding}, the spatial embedding selects a single row by the total degree of the road, the number of roads that meet it at either endpoint, i.e., its degree in the controller graph $L (G)$. It thus no longer tells how these roads are split between the two head junctions of the road.

\item \emph{No temporal embedding:} $E'_{\mathrm{T}}$ is dropped from \eqref{eq:input-embedding}, so temporal order enters only through the causal mask of \eqref{eq:mask}.

\end{itemize}

All ablations still improve on the baseline, by $12.4 \%$ to $21.9 \%$ in return, but they fall into two groups. Removing the relational bias or the spatial embedding lowers the return by $7.9 \pm 0.3 \%$ and $8.5 \pm 0.4 \%$ against the full model and multiplies the delay by about $4.7$ and $5.0$, to 1,742 and 1,862 veh\textperiodcentered{}h. Both variants leave vehicles in gridlocked queues, $159$ and $87$ per day, and have far more closing proposals rejected, $670$ and $560$ per day against $161$. Their greedy policies stay below that of the full model at every evaluation from update $30$ on. Without the relational bias, the encoder also forecasts worst, with a validation MSE $7.9 \%$ above that of the full model, whereas without the spatial embedding, it forecasts as well as the full model (Fig.~\ref{fig:ablations} (b)).

The other three ablations, collapsed relations, the undirected spatial embedding, and no temporal embedding, end within about $1 \%$ of the full model in return, at $- 1.0 \pm 0.1 \%$, $- 0.8 \pm 0.2 \%$, and $- 1.0 \pm 0.3 \%$, but their final policies incur $1.5$, $1.4$, and $1.5$ times its delay, $547$, $516$, and $552$ against $374$ veh\textperiodcentered{}h per day. Across the evaluation days, these differences far exceed their standard errors, but with one training run per variant, they also depend on when training stopped: from update $30$ on, the greedy return of the full model varies by $680$ veh\textperiodcentered{}h, about $4 \%$, more than these variants lose at the end. Over these earlier evaluations, the three variants differ. Without the temporal embedding, the greedy return lies below that of the full model at $6$ of the $8$ evaluations, by $1.0 \%$ on average, and the forecasting error lies above it at every validation from step 1,000 on, ending $2.1 \%$ higher (Fig.~\ref{fig:ablations} (b)). With collapsed relations, by contrast, the greedy return exceeds that of the full model at $7$ of the $8$ evaluations, by $1.2 \%$ on average, and with the undirected spatial embedding at $5$ of them, and both variants forecast as well as the full model, within about $0.5 \%$. The final evaluation thus suggests that all three components reduce delay, but only for the temporal embedding do the earlier evaluations and pretraining agree.

%% file: sections/discussion.tex
\section{Discussion}\label{sec:discussion}
Pretraining likely helps because the forecasting task asks the encoder for what both the actor and the critic need: how the congestion, density, and velocity of every road will evolve over the next hour, given the lane allocations and speed limits that take effect in between. The forecasts of a road depend on the roads that feed and drain it, so to forecast well, the encoder must represent how traffic propagates through the network and how it responds when a direction is closed or slowed, and it forecasts far better than persistence, with errors $59 \%$ to $77 \%$ below it. In fine-tuning, by contrast, an action reaches the reward only indirectly and late: closing a direction diverts the traffic headed for it at once, but its lanes open in the other direction only once clearing completes, queues build and spill back over several decision steps, and with $\alpha_{\mathrm{S}} = 0$, the mixed reward of a road is dominated by the actions of other roads. Learning the same representation from this signal alone is slow, as the untrained encoder shows: it needed about twice as many updates to surpass the baseline and left more than twice the delay after $100$ updates. The initial dip of the training episodes is alike in both runs (Fig.~\ref{fig:pretraining} (b)) and says nothing about the encoders: it is a transient of the staggered start, during which the actor's encoder is frozen or has barely been trained. Both runs used the same fine-tuning schedule, which was chosen for a pretrained encoder: it freezes the encoders during the first updates and then trains them at a lower learning rate than the heads. A schedule tuned for random initialization might narrow the gap.

The learned policy beats the baseline by reallocating lanes in tidal flow. During each rush hour, it assigns most lanes to the peak direction and makes about two thirds of the roads one-way in that direction, and around midday, when demand is balanced, it returns to two-way roads. It thereby raises the capacity in the peak direction by more than half, to $155$ and $159$ of the $200$ lanes instead of $100$, which keeps queues from forming at the roads where trips enter the network and on the roads near the two zones, and from spilling back into gridlock. The policy learned this pattern from the reward alone; neither the behavior policy of pretraining nor the initial configuration suggests it. Rerouting makes it workable: a vehicle whose next direction closes takes another route, which strong connectivity guarantees to exist (Proposition~\ref{prop:connectivity}), and the mean travel time still falls by a third despite such detours. One-way roads appear to be essential to this gain. Half of the roads of our grid have only two lanes, which a controller that keeps every road two-way cannot reallocate at all, and on the other roads, such a controller can open at most $L_{r} - 1$ lanes in one direction, at most $140$ of the $200$ lanes in all. Within these bounds, the load balancer and the two-way restrictions of the policy leave six to eight times the delay of the learned policy, and none of them prevents gridlock. A policy trained for two-way roads might do better than the restricted one, but it would face the same bounds.

The learned speed limits, by contrast, cost return. Holding them nominal at test time removes more than half of the remaining delay, and applied to the initial configuration, they increase its delay. The delay charges every vehicle for the time it loses below the nominal speed limit, so a lower speed limit costs delay on its own road and pays off only if it prevents more delay elsewhere. In our simulator, moreover, a lane discharges at a rate that a lower speed limit can only reduce, and its capacity does not drop under congestion, so the mechanisms by which VSLs delay the onset of congestion and dampen shock waves on freeways \cite{ext:lu2014vsl} have no counterpart. The policy keeps its speed limits close to nominal, but even the small reductions in Fig.~\ref{fig:day} (e) cost more than they save. Why fine-tuning did not remove them is not clear from our runs; possibly, the delay that a speed limit causes on its own road enters the mixed reward of that road only with the weight $\alpha_{\mathrm{G}} / R$ (Section~\ref{sec:exp:fine-tuning:reward}), which is small against the variation contributed by all other roads. Whether learned speed limits can contribute to joint control thus remains open; answering it would need a simulator with the congestion mechanisms on which VSLs act.

Action resolution, finally, rarely has to intervene during the evening rush hour: between 16:00 and 19:30, only $4$ proposals are rejected, although the policy keeps $35$ to $40$ roads one-way around the peak, so it largely proposes closings that keep the network strongly connected. Most rejections occur when it switches many roads to one-way at once, at the onset of the morning rush hour, and at night, when lane allocations hardly affect the reward and the policy keeps up to $34$ of the $60$ roads one-way and proposes conflicting closings at no cost. Since every road of our grid has at least two lanes, a rejected proposal is always relaxed to the nearest allocation that keeps one lane open, which costs the policy little and may explain why it does not avoid such proposals entirely.

Among the components of attention, the relational bias matters most. Without it, attention is global but blind to which roads are adjacent: every road is related to every other in the same way, and the model must infer which roads exchange traffic from correlations in their features. This variant forecasts worst of all, and its policy incurs almost five times the delay of the full model and has four times as many closing proposals rejected, which suggests that without relations, the actor can tell neither where shed traffic goes nor which closings conflict. Telling apart the directions in which adjacent roads meet, by contrast, helped less clearly: collapsing the four adjacency relations into one left the forecasting error unchanged, and although the resulting final policy incurs $1.5$ times the delay of the full model, this variant did better than the full model at most earlier evaluations. A possible reason is the uniform orientation of our grid, in which every road runs forward eastward or southward. The forward and backward features of adjacent roads then refer to the same compass directions, so the model may learn from their traffic which directions of adjacent roads queue together, and thus how spillback propagates through a shared junction, without being told. Randomizing the reference ordering of the endpoints of every road in each episode would remove this shortcut and leave the endpoint relations as the only indication of how the directions of adjacent roads connect, so training both the full model and the variant with collapsed relations under such randomization and comparing them would measure the contribution of the endpoint relations. Since the orientation of a real network is arbitrary, such randomization is also a natural step toward transfer to other networks, and irregular networks, on which the hop distance says less about how two roads interact, would test the endpoint relations further.

The spatial embedding matters for control but not for forecasting: without it, the forecasting error is unchanged, but the policy incurs five times the delay of the full model and has $3.5$ times as many closing proposals rejected. The junction degrees carry information that attention cannot easily recover, since attention averages over the attended tokens with weights that sum to one and therefore cannot directly count the roads at a junction, even with the relational bias; Graphormer likewise complements attention with a degree-based centrality encoding \cite{ext:ying2021graphormer}. For forecasting, the number of roads at a junction matters little once the traffic on them is observed, but for control, it bears on which closings keep the network strongly connected: a junction with few roads has few alternatives for entering and leaving it, so it is easily cut off. The evidence on how the adjacent roads are split between the two ends of a road is mixed as well: with the undirected embedding, the final policy incurs $1.4$ times the delay of the full model, yet the variant led the full model at $5$ of the $8$ earlier evaluations. On our grid, the total degree already identifies whether a road touches a corner, lies on the boundary, leads into the interior, or lies within it, so the undirected embedding loses only which of its ends lies toward the boundary. The temporal embedding, finally, shows the most consistent benefit of the three, if a small one: without it, the forecasting error is higher throughout pretraining, the greedy return is lower at most evaluations, and the final policy incurs $1.5$ times the delay of the full model and has $2.4$ times as many closing proposals rejected. The benefit is possibly small because the causal mask already conveys temporal order: tokens at different positions attend to different numbers of positions, from which the stacked layers can infer how recent a position is, and the traffic of a road changes gradually over a window, so the order of its time steps can largely be read from their content. Repeating this ablation with a mask without causality, that is, comparing the model with and without the temporal embedding when every token of both may attend to all valid positions of its window, would isolate the temporal embedding. Since every window ends at its decision step and only its last position is read out, such a mask would use no information from the future, and without the temporal embedding, the model would then see the earlier time steps of a window, apart from the last one at which it is read out, as an unordered set, so this comparison would measure all the temporal order that the embedding provides.

Our evaluation has several limitations. It uses a single synthetic network with one demand pattern, a $6 \times 6$ grid with two rush hours between opposite corners, which suits the tidal pattern that the policy learned. Each variant was trained once, so differences of about $1 \%$ in return between variants, even with $1.5$ times the delay, cannot be firmly attributed to the architecture; several training seeds per variant would be needed. Of our two baselines, the fixed configuration ignores the traffic, and the load balancer responds to it but keeps every road two-way; a schedule that makes roads one-way toward the commercial zone in the morning and away from it in the evening would show how much of the gain requires a policy that responds to the observed traffic. All vehicles are homogeneous CAVs that comply fully and route themselves with knowledge of the current traffic, so mixed traffic and partial compliance are not modeled. The mesoscopic simulator has no signals or turning conflicts at junctions and no capacity drop, which limits what VSLs can achieve, and its gridlock safeguard removes vehicles that would otherwise keep accruing delay, which flatters the baseline. Finally, attention over all $W_{\max} R$ road--time tokens grows quadratically with the number of roads: our $60$ roads give $960$ tokens per sample, and a city network with thousands of roads would need sparser attention, for example restricted to a hop neighborhood or factorized over space and time. In our implementation, the relational bias alone doubles the time of a pretraining step, $0.21$ against $0.11$ s. Since no parameter of the model depends on the network (Section~\ref{sec:arch}), the most direct extension is to train on several networks and test transfer to unseen ones, including real road networks.

%% file: sections/conclusion.tex
\section{Conclusion}\label{sec:conclusion}
We formulated network-wide dynamic lane allocation and directional speed limit control with one-way roads, explicit lane clearing, and a strong-connectivity requirement on the navigable graph, and gave an action-resolution algorithm that keeps every junction reachable from every other at every time step, for any admissible actions and however long clearing takes. All road controllers share a single GERASTSA transformer, pretrained on action-conditioned traffic forecasting and fine-tuned with per-road PPO under a mixed reward. On a $6 \times 6$ grid with commuter rush hours, the learned policy reallocated lanes in tidal flow, making about two thirds of the roads one-way in the peak direction during each rush hour. It reduced the daily delay by $90 \%$ and the mean travel time by $32 \%$ against holding the initial configuration with nominal speed limits, and the delay by $85 \%$ against a tuned load balancer that keeps every road two-way. Restricted to two-way roads at test time, the policy gained only about as much as the load balancer, which suggests that its gain rests on one-way roads, whereas its learned speed limits cost return: holding them nominal more than halved its delay. Pretraining roughly halved the number of updates needed to surpass the fixed configuration and left less than half the delay of fine-tuning from random initialization. The relational bias and the spatial embedding mattered most, as removing either multiplied the delay by about five. Removing the endpoint relations, the direction of the spatial embedding, or the temporal embedding left the final policies with $1.4$ to $1.5$ times the delay, an effect that the earlier evaluations and pretraining corroborate only for the temporal embedding. The most promising direction for future work is to train the model on several networks, with the orientation of their roads randomized in each episode, and evaluate it on unseen ones, real road networks included, which its network-independent parameters permit. Two ablations, each repeated under a changed condition and with several training seeds, would measure the contributions of the endpoint relations and the temporal embedding: collapsing the endpoint relations with the orientation of the roads randomized in each episode, and removing the temporal embedding with a mask without causality. A schedule of one-way roads as a further baseline would show how much of the gain requires a learned policy, and a simulator with capacity drop whether learned speed limits can contribute at all.

%% file: sections/acknowledgments.tex
\section*{Acknowledgments}
The authors thank Kaggle for free access to GPUs, which supported the experiments of this work.

During the preparation of this work, the authors used Claude and ChatGPT to improve manuscript clarity, check notational consistency, and assist with programming. The authors reviewed and edited all AI-assisted content and take full responsibility for the content of this publication.